\documentclass[%
reprint,
superscriptaddress,
showpacs,preprintnumbers,
amsmath,amssymb,
prb,
]{revtex4-1}
\usepackage{graphicx}
\usepackage{dcolumn}
\usepackage{xfrac}
\usepackage{bm}
\usepackage{color}
\usepackage[export]{adjustbox}
\usepackage{xr-hyper}
\usepackage{hyperref}
\usepackage{tabularx}

\newcommand{\EZA}{EuZn$_2$As$_2$}

\newcommand{\ECA}{EuCd$_2$As$_2$}
\newcommand{\ECP}{EuCd$_2$P$_2$}
\newcommand{\C}{$^\circ$C}
\begin{document}

\title{Anisotropy of the magnetic and transport properties in EuZn$_2$As$_2$}

\author{Zhi-Cheng~Wang}
\thanks{The authors contributed equally to this work.}
\affiliation{Department of Physics, Boston College, Chestnut Hill, MA 02467, USA}

\author{Emily~Been}
\thanks{The authors contributed equally to this work.}
\affiliation{Stanford Institute for Materials and Energy Sciences, SLAC National Accelerator Laboratory, 2575 Sand Hill Road, Menlo Park, CA 94025.}
\affiliation{Department of Physics, Stanford University, Stanford, CA 94305.}

\author{Jonathan~Gaudet}
\affiliation{NIST Center for Neutron Research, National Institute of Standards and Technology, Gaithersburg, Maryland 20899, USA}
\affiliation{Department of Materials Science and Eng., University of Maryland, College Park, MD 20742-2115}

\author{Gadeer~Matook~A.~Alqasseri} 
\affiliation{Department of Physics and Astrophysics, Howard University, Washington DC, 20059.}
\affiliation{IBM-Howard Quantum Center, Howard University, Washington DC, 20059.}

\author{Kyle~Fruhling}
\author{Xiaohan~Yao}
\affiliation{Department of Physics, Boston College, Chestnut Hill, MA 02467, USA}

\author{Uwe~Stuhr}
\affiliation{Laboratory for Neutron Scattering and Imaging, Paul Scherrer Institut, 5232 Villigen-PSI, Switzerland}

\author{Qinqing~Zhu} 
\author{Zhi~Ren}
\affiliation{School of Science, Westlake University, 18 Shilongshan Road, Hangzhou 310064, PR China.}
\affiliation{Institute of Natural Sciences, Westlake Institute for Advanced Study, 18 Shilongshan Road, Hangzhou 310064, PR China.}

\author{Yi~Cui}
\affiliation{Department of Materials Science and Engineering, Stanford University, Stanford, CA 94305.}

\author{Chunjing~Jia} 
\affiliation{Stanford Institute for Materials and Energy Sciences, SLAC National Accelerator Laboratory, 2575 Sand Hill Road, Menlo Park, CA 94025.}

\author{Brian~Moritz}
\affiliation{Stanford Institute for Materials and Energy Sciences, SLAC National Accelerator Laboratory, 2575 Sand Hill Road, Menlo Park, CA 94025.}

\author{Sugata~Chowdhury} 
\affiliation{Department of Physics and Astrophysics, Howard University, Washington DC, 20059.}
\affiliation{IBM-Howard Quantum Center, Howard University, Washington DC, 20059.}

\author{Thomas~Devereaux}
\affiliation{Stanford Institute for Materials and Energy Sciences, SLAC National Accelerator Laboratory, 2575 Sand Hill Road, Menlo Park, CA 94025.}
\affiliation{Department of Materials Science and Engineering, Stanford University, Stanford, CA 94305.}

\author{Fazel~Tafti}
\email{fazel.tafti@bc.edu}
\affiliation{Department of Physics, Boston College, Chestnut Hill, MA 02467, USA}

\date{\today}

\begin{abstract}
Several recent studies have shown that the anisotropy in the magnetic structure of \ECA\ plays a significant role in stabilizing the Weyl nodes.
To investigate the relationship between magnetic anisotropy and Weyl physics, we present a comparative study between \EZA\ and \ECA\ that are isostructural but with different magnetic anisotropy.
We performed structural analysis, electronic transport, and magnetization experiments on millimeter-sized single crystals of \EZA, and compared the results to those of \ECA. 
By combining the first principle calculations and neutron diffraction experiment, we identify the magnetic ground state of \EZA\ as A-type antiferromagnetic order with a transition temperature ($T_\mathrm{N}$ = 19.6 K) twice that of \ECA. 
Like \ECA, the negative magnetoresistance of \EZA\ is observed after suppressing the resistivity peak at $T_\mathrm{N}$ with increasing fields. However, the anisotropy in both transport and magnetization are much reduced in \EZA. 
The difference could be ascribed to the weaker spin-orbit coupling, more localized $d$-orbitals, and a larger contribution from the Eu $s$-orbitals in the zinc compound, as suggested by the electronic band calculations. The same band structure effect could be also responsible for the observation of a smaller non-linear anomalous Hall effect in \EZA\ compared to \ECA.
\end{abstract}

\maketitle


\section{\label{sec:intro}Introduction}
Recent observations of anisotropic magnetoresistance, spin-fluctuation-induced Dirac nodes, and non-linear anomalous Hall effect in \ECA\ have made this material an interesting candidate to study the interplay between topology and magnetism~\cite{EuCd2As2_PRB_2016,AHE_EuCd2As2,SA_EuCd2As2_2019,EuCd2As2_PRB_2018,PRB_EuCd2As2_2019_JRSoh,PRB_EuCd2As2_2019_LLWang,Ma_EuCd2As2_2020,soh_resonant_2020}.
\ECA\ undergoes an A-type antiferromagnetic (AFM) order at 9.2~K with considerable anisotropy between the in-plane and out-of-plane magnetic susceptibilities~\cite{Ma_EuCd2As2_2020}. 
Its resistivity shows a peak near the N\'{e}el temperature ($T_\mathrm{N}$) which is suppressed in an external magnetic field, and thus may be related to the fluctuations of the AFM order~\cite{LaMnO,Tl2Mn2O7,FeCr2S4,Eu14MnSb11,Yb14MnSb11,EuIn2P2,EuIn2As2}.
Its Hall resistivity also shows a peak that is non-linear in either the magnetic field or magnetization, hence the name non-linear anomalous Hall effect (NLAHE)~\cite{AHE_EuCd2As2}.
To bring the resistivity peak and NLAHE to higher temperatures, it is necessary to increase the temperature scale of the AFM order and its fluctuations.
Here we present \EZA\ as an analogue of \ECA\ but with a $T_\mathrm{N}$ twice as high, possibly due to the weaker spin-orbit coupling (SOC) and more localized $d$-orbitals in the zinc compound. 
We show that (a) both the resistivity peak and NLAHE are shifted to higher temperatures in \EZA\ compared to \ECA, and (b) the anisotropy in both transport and magnetization are reduced in the zinc compound. 
We present a comprehensive study of the crystal structure, magnetic susceptibility, heat capacity, anomalous Hall effect of \EZA\ and \ECA, and theoretical electronic structure to understand the factors that control the anisotropy of physical properties and their temperature scales in these materials.

\section{\label{sec:exp}Methods}

\textit{Crystal Growth.} Single crystals of \EZA~were grown in Sn flux, by using sublimed ingots of europium (99.9\%), zinc powder (99.9\%), arsenic powder (99.99\%), and tin shots (99.999\%) as the starting materials. 
The elements were mixed in a mole ratio Eu:Zn:As:Sn = 1:1:1:8. 
The excess of Eu prohibits the formation of Zn$_3$As$_2$ impurity. 
The mixture was loaded into an alumina crucible inside an evacuated quartz ampule and slowly heated to 1100~\C, held for 24~h, cooled to 900~\C\ at 3~\C/h, cooled to 600~\C\ at 5~\C/h, and finally centrifuged to remove the flux. 
The crystals grow as millimeter-sized hexagonal prisms with metallic luster, and are stable in air.

\textit{Powder X-ray Diffraction.} A few crystals were ground for the powder X-ray diffraction using a Bruker D8 ECO instrument equipped with 40 keV copper source and a 1D LYNXEYE XE detector. 
The FullProf suite was used for the structural refinements~\cite{rodriguez-carvajal_recent_1993}.

\textit{Resistivity, Heat capacity, and Magnetization Measurements.} Transport data were collected with a standard four-probe technique using a Quantum Design Physical Property Measurement System (PPMS) Dynacool with a high-resolution rotator option. 
The heat capacity data were measured using the PPMS with a relaxation time method. 
Crystals with clear facets were selected to measure DC magnetization using a Quantum Design Magnetic Property Measurement System (MPMS-3).

\textit{Neutron Diffraction.} A single crystal of \EZA\ was mounted on an Al plate with the $\vec{b}$ axis vertical to the scattering plane so the Bragg scattering within the ($h0l$) plane could be probed. 
The experiment was performed with the thermal neutron triple-axis spectrometer EIGER at the SINQ spallation neutron source at the Paul Scherrer Institute (PSI). 
The energy of the incident neutrons was 14.7~meV with a resolution of $\sim$0.66~meV. 
A fully open collimation configuration was used to maximize the neutron flux as well as PG filters to eliminate high-order Bragg reflections.
Representational analysis of magnetism was performed using SARAh~\cite{SARAh}.

\textit{First-principles Calculations.}\label{methods-theory} The total energies and band structures were calculated using the projector-augmented wave (PAW) version \cite{VASP-PAW_Kresse_Joubert_PRB1999} of density functional theory (DFT) with the Perdew-Burke-Ernzerhof (PBE) exchange correlation functional \cite{PBE}, as implemented in Vienna \textit{ab initio} simulation package (VASP) \cite{VASP_Kresse_Furthmuller_CompMatSci1996,VASP_Kresse_Furthmuller_PRB1996,VASP_Kresse_Hafner_PRB1993,VASP_Kresse_Hafner_PRB1994}. 
The pseudopotential for Eu had $4f$ electrons explicitly in the valence, therefore we added a Hubbard $U=5$~eV and spin-orbit coupling (SOC) to correctly describe their localization \cite{PRB_EuCd2As2_2019_LLWang,EuIn2As2_axionInsulator}. 
These GGA+$U$+SOC calculations were performed using the rotationally invariant method by Dudarev \emph{et al}. \cite{LDAUTYPE2}.
The self-consistent field (SCF) calculations used a $k$-point Monkhorst-Pack grid of $8\times 16\times 16$ for the 2-Eu supercell, $8\times 16\times 8$ for the 4-Eu supercell, $16\times16\times4$ for 3-Eu supercell, and $5\times5\times9$ for the 9-Eu supercell~\cite{suppmatt}. 

\section{\label{results}Results and Discussions}

\begin{figure}
	\includegraphics[width=0.45\textwidth]{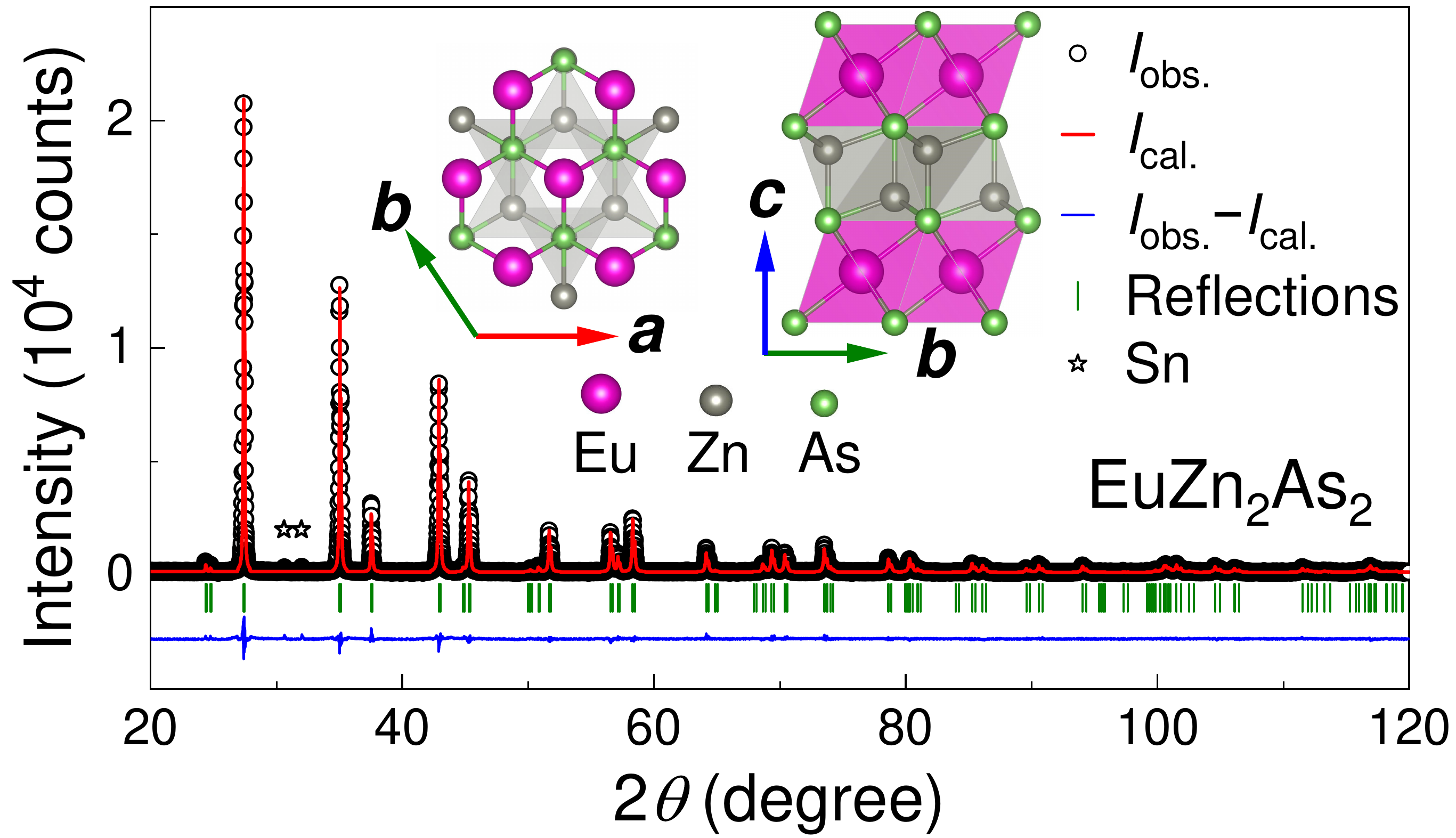}
	\caption{\label{fig:XRD}
		Powder X-ray diffraction of \EZA (black circles), Rietveld refinement in space group $P\bar{3}m1$ (red lines), fit residual curve (blue lines), and $hkl$ indices (green ticks) are shown. 
		The black stars show Bragg peaks from the Sn flux.
		Inset shows the crystal structure viewed from [001] and [100] directions. 
		The Eu, Zn, and As atoms are shown as magenta, green, and gray balls, respectively.
		The structual refinement of \ECA\ is presented in the supplemental Fig.~S1.
	}
\end{figure}

\subsection{Structural analysis} 
The powder X-ray diffraction pattern of \EZA\ in Fig.~\ref{fig:XRD} is refined in the trigonal space group $P\bar{3}m1$ (\#164).
The structural parameters are listed in Table~\ref{tab:tab1} and compared to those of the sister compound \ECA. 
Inset of Fig.~\ref{fig:XRD} shows that the structure comprises alternating layers of edge-sharing ZnAs$_4$ tetrahedra and edge-sharing EuAs$_6$ octahedra.
Table~\ref{tab:tab1} shows that all bond lengths are slightly shorter and bond angles are slightly wider in the Zn compound.
Despite subtle changes of crystal structure between \EZA\ and \ECA, we will show in Section~\ref{subsec:dft} that the electronic structure changes visibly and leads to changes of electronic and magnetic anisotropy between the two compounds.

\begin{table}
	\caption{\label{tab:tab1} Crystallographic data and Rietveld refinement parameters for \EZA\ and \ECA\ in the space group $P\bar{3}m1$.
	The Wyckoff sites are Eu~$1a$ (0,0,0), Zn/Cd~$2d$ (1/3,1/3,$z$), and As~$2d$ (1/3,2/3,$z$) with full occupancy.}
	\begin{ruledtabular}
		\begin{tabular}{lll}			
			Material & \EZA\ & \ECA\ \\
			\hline
			Lattice parameters & & \\
			$a$ (\AA) & 4.21118(3)& 4.44016(4) \\ 
			$c$ (\AA) & 7.18114(6) & 7.32779(9) \\
			$V$ (\AA$^3$)& 110.2888(24) & 125.1125(38) \\
			$c/a$ & 1.705 & 1.650 \\
			$Z$ & 1 & 1 \\
			\hline
			Coordinates ($z$)&&\\
			Zn/Cd & 0.62859(16) & 0.63342(21) \\
			As & 0.26743(16) & 0.24593(29) \\
			\hline
			Debye-Waller factors $B_\mathrm{iso}$ &&\\
			Eu (\AA$^2$) & 1.183(28) & 1.820(90) \\
			Zn/Cd (\AA$^2$) & 1.738(43) & 1.813(86) \\
			As (\AA$^2$) & 1.163(32) & 1.840(83) \\
			\hline
			$R$-factors &&\\
			$R_\mathrm{p}$ & 6.95 & 8.22 \\
			$R_\mathrm{exp}$ & 6.25 & 6.59 \\
			$\chi^2$ & 2.25 & 2.47\\
			\hline
			Bond distances (\AA) & & \\
			Eu-As ($\times6$) & 3.0983(8) & 3.1336(13) \\
			As-Zn/Cd ($\times3$) & 2.5434(5) & 2.7117(9) \\
			As-Zn/Cd ($\times1$) & 2.5935(17) & 2.8395(3) \\
			\hline
			Bond angles ($^\circ$) & & \\
			As-Zn/Cd-As & 111.76(4) & 109.91(6) \\
		\end{tabular}
	\end{ruledtabular}
\end{table}

\begin{figure*}
	\includegraphics[width=\textwidth]{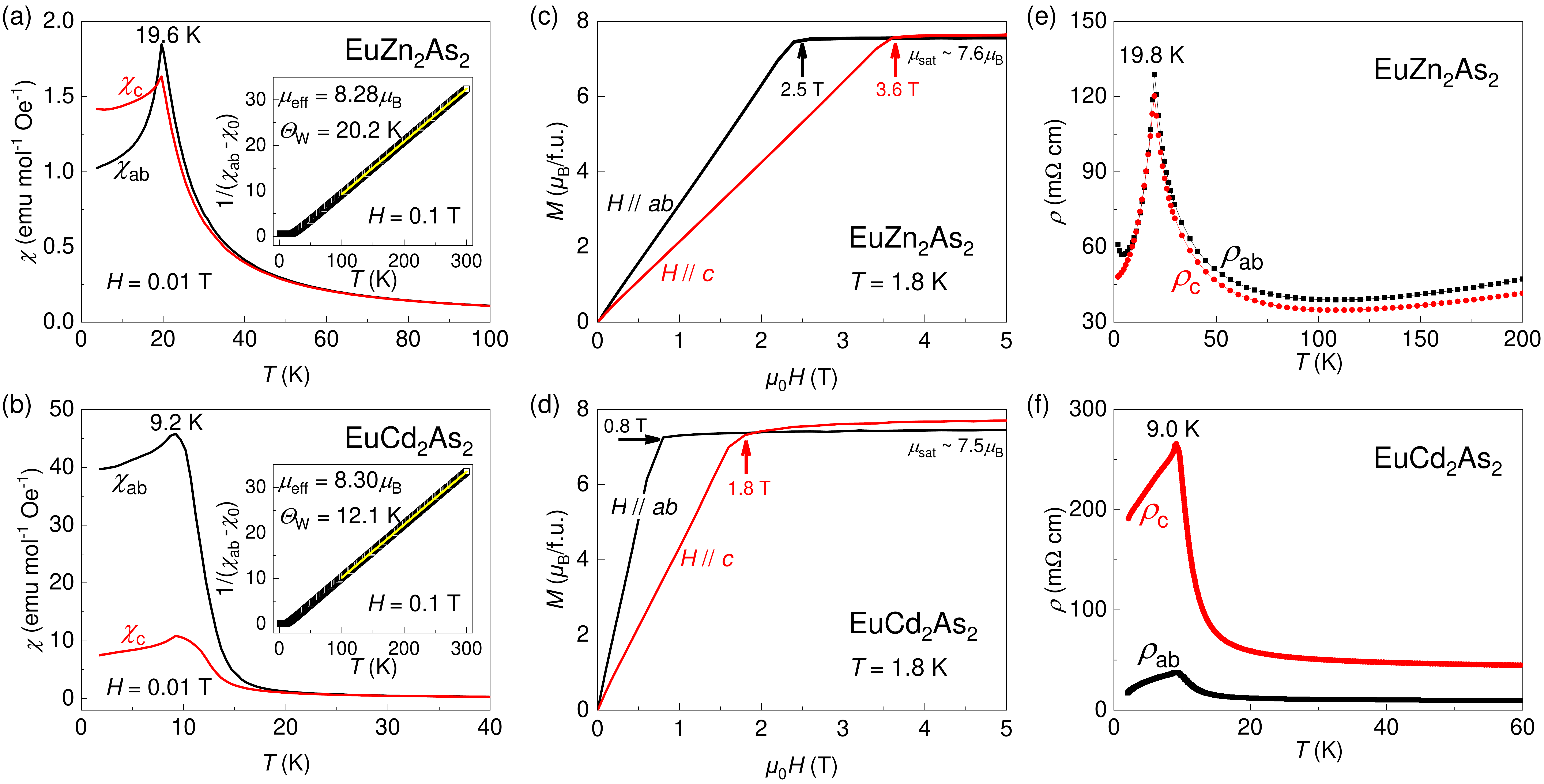}
	\caption{\label{fig:ANISOTROPY}
		Temperature dependence of anisotropic magnetic susceptibility with in-plane ($\chi_\mathrm{ab}$) and out-of-plane fields ($\chi_\mathrm{c}$) are plotted for (a) \EZA~and (b) \ECA. The insets show the Curie-Weiss analysis. Magnetization curves as a function of both in-plane (black) and out-of-plane (red) fields are plotted for (c) \EZA~and (d) \ECA. Temperature dependence of resistivity is plotted with in-plane (black) and out-of-plane (red) current for (e) \EZA\ and (f) \ECA. Data in panel (f) is from reference~\onlinecite{EuCd2As2_PRB_2016}. More information is provided in the supplemental Fig.~S2.
	}
\end{figure*}

\subsection{\label{subsec:Mag}Magnetic and Transport Anisotropy} 
\EZA\ and \ECA\ order at $T_\textrm{N}=19.6$~K and $9.2$~K, respectively.
The magnetic transition is marked by a peak in the susceptibility data under both in-plane and out-of-plane fields ($\chi_{ab}$ and $\chi_c$ in Figs.~\ref{fig:ANISOTROPY}a,b).
The transition is considered antiferromagnetic (AFM) based on (i) absence of splitting between the zero-field-cooled (ZFC) and field-cooled (FC) susceptibility data (supplemental Fig.~S2) and (ii) absence of hysteresis in $M(H)$ curves (Figs.~\ref{fig:ANISOTROPY}c,d).
The Curie-Weiss fits in the insets of Figs.~\ref{fig:ANISOTROPY}a,b show effective moments of approximately 8.3~$\mu_\mathrm{B}$ in both compounds, slightly larger than the theoretical value of 8~$\mu_\mathrm{B}$ for Eu$^{2+}$. 
The saturated moments at high fields in Figs.~\ref{fig:ANISOTROPY}c,d are also slightly larger than the theoretical value of 7~$\mu_\mathrm{B}$ for Eu$^{2+}$.
Both observations are due to a known effect, namely the polarization of $d$-orbitals induced by the large $f$-moments in either Eu$^{2+}$ or Gd$^{3+}$ with $4f^7$ configuration~\cite{PRL_fdcouple,PRB_fdcouple}.

The Curie-Weiss fits (insets of Figs.~\ref{fig:ANISOTROPY}a,b and supplemental Fig.~S3) yield positive Weiss temperatures ($\Theta_\textrm{W}>0$) indicative of ferromagnetic (FM) correlations in both compounds.
The presence of FM correlations despite AFM ordering implies either A-type AFM (inter-layer AFM and intra-layer FM coupling) or C-type AFM order (inter-layer FM and intra-layer AFM coupling).
Recent resonant elastic x-ray scattering (REXS) experiments have confirmed the A-type AFM order in \ECA~\cite{EuCd2As2_PRB_2018}.
Here, we investigate the magnetic structure of \EZA.

There is a visible difference in the anisotropy of the magnetic susceptibility between \EZA\ and \ECA.
Figures~\ref{fig:ANISOTROPY}a,b show that the ratio $\chi_{ab}/\chi_c$ is near one in \EZA\ and near four in \ECA. 
Figures.~\ref{fig:ANISOTROPY}c,d show that the ratio of saturation fields with $(H\|c)/(H\|ab)$ is $(3.6~\text{T})/(2.5~\text{T})=1.4$ in \EZA\ versus $(1.8~\text{T})/(0.8~\text{T})=2.2$ in \ECA.
These observations can be interpreted in two ways.
Either, the magnetic ordering in \EZA\ is C-type unlike A-type in \ECA, or it is A-type but with smaller anisotropy.
We will show the latter to be correct using neutron scattering and first-principle calculations.

The smaller magnetic anisotropy of \EZA\ compared to \ECA\ is also reflected in the resistivity data in Figs.~\ref{fig:ANISOTROPY}e,f.
The zero-field resistivity of \EZA\ is isotropic (Fig.~\ref{fig:ANISOTROPY}e) whether the electric current is applied out-of-plane ($J\|c\to\rho_c$) or in-plane ($J\|ab\to\rho_{ab}$).
In contrast, the resistivity of \ECA\ is anisotropic with $\rho_c/\rho_{ab}\approx 5$ from room to low temperatures (Fig.~\ref{fig:ANISOTROPY}f)~\cite{EuCd2As2_PRB_2016}.
In both cases, $\rho(T)$ shows a peak near $T_N$ as it increases in the region of magnetic fluctuations $T_N<T<3T_N$ and decreases at $T<T_N$.
Also, in both cases, the room temperature resistivity is more than 20~m$\Omega$cm, which is 20 times larger than the localization (Ioffe-Regel) limit~\cite{ioffe_non-crystalline_1960}.
Nonetheless, $\rho(T)$ shows a mild decrease with temperature.
This so-called bad metal behavior is observed in conductors with strong correlations and/or incoherent scattering due to spin fluctuations~\cite{PRL_badmetal,AM_EuCd2P2_2021}.

The observation of a FM Weiss temperature despite AFM ordering in both \EZA\ and \ECA\ suggests a competition between FM and AFM correlations.
We used first-principles calculations to analyze the total energies of various magnetic structures. 
All combinations of in-plane versus out-of-plane spins, FM or type-A, -C, -G AFM, and two types of 120$^{\circ}$ rotational cells were considered. 
Details of these calculations can be found in Section~IV of the Supplemental Material (SM)~\cite{suppmatt}.
These calculations predict that both \ECA\ and \EZA\ have an A-type AFM ground-state (Fig. S6), which for \ECA\ is consistent with REXS experiments.
A striking observation from this analysis is the small energy differences between all magnetic configurations.
In \EZA, the largest energy difference is about 5~meV between the G-type and A-type AFM states with the latter being the ground-state.
In \ECA, the energy difference is even smaller, approximately 2~meV between the G-type and A-type AFM states.
Such small energy differences approach the computational resolution, i.e. both C-type and A-type AFM states could be the ground-state of either compound.
The experimental implication is that the magnetic ground-state of these materials can be easily manipulated by external pressure or strain, which creates an exciting platform for tuning a topological band structure by tuning the magnetic structure~\cite{PRB_EuCd2As2_2019_LLWang}.

\subsection{Neutron Diffraction} 
\begin{figure}
	\includegraphics[width=0.45\textwidth]{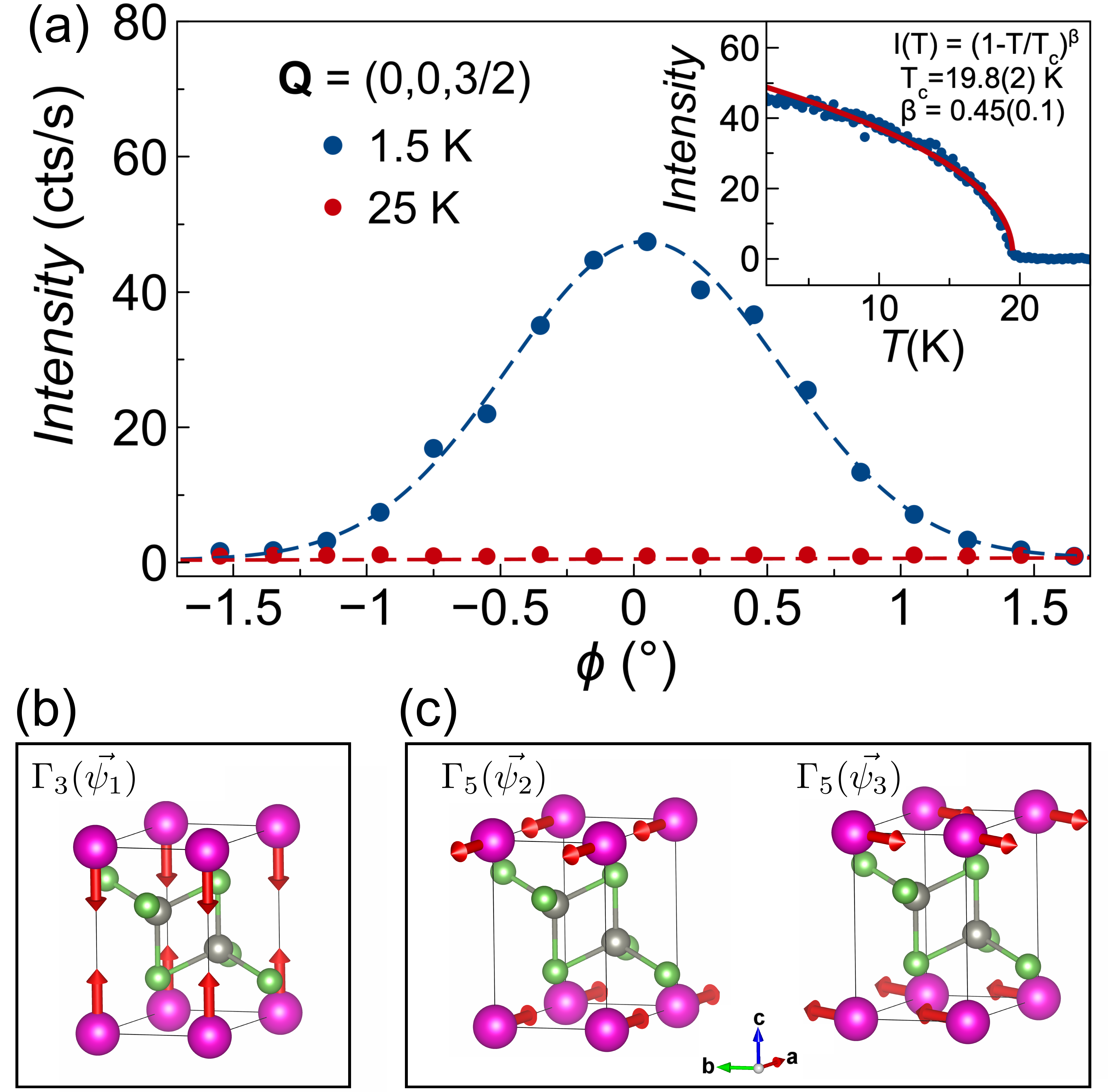}
	\caption{ (a) {Rocking scan measured at $\vec{Q}$~=~(0,0,3/2) and collected for both $T$~=~25~K and $T$~=~1.5~K. The top right inset shows the order parameter measurement, which was also acquired at $\vec{Q}~=~(0,0,3/2)$. The real part of the basis vectors resulting from the representation symmetry analysis of space group 164 with a $\vec{k}$~=~(0,0,1/2) ordering vector is presented in (b) and (c). Panel b (c) shows the $\vec{\psi_1}$ ($\vec{\psi_2}$ and $\vec{\psi_3}$) basis vector(s) corresponding to the $\Gamma_3$ ($\Gamma_5$) manifold where strictly out-of-plane (in-plane) spin component is allowed. Eu, Zn, As atoms are shown as magenta, green, gray balls, respectively.}
	\label{fig:Neutron}
}
\end{figure}
Neutron diffraction can unambiguously distinguish between the A- and C-type magnetic ordering in \EZA.
The magnetic Bragg peaks ($\vec{Q}_{mag}$) of the A- and C-type orders are respectively indexed by $\vec{k}$~=~(0,0,1/2) and (1/2,0,0) ordering vectors ($\vec{Q}_{mag}~=~\vec{Q}_{nuc} \pm \vec{k}$ where $\vec{Q}_{nuc}$ is the position of a nuclear Bragg peak).
Thus, we collected neutron diffraction scans along the (H00) and (00L) directions within the first Brillouin zone of various nuclear zone centers at 1.5~K, well below $T_\mathrm{N}=19.6$~K. 
Eleven Bragg peaks were found that could be indexed by a $\vec{k}$~=~(0,0,1/2) ordering vector (see rocking scans for $\vec{Q}$=(0,0,3/2) plotted in Fig.~\ref{fig:Neutron}a as an example), while no $\vec{k}$~=~(1/2,0,0) Bragg peaks were observed.
Also, no $\vec{k}$~=~(0,0,0) peaks were observed, ruling out FM configurations consistent with the absence of $M(H)$ hysteresis in Fig.~\ref{fig:ANISOTROPY}c.
Thus, neutron scattering confirms A-type AFM order predicted by DFT.

Note that within the A-type order, symmetry analysis tells us that the spins can point either out-of-plane or in-plane (Figs.~\ref{fig:Neutron}b,c), corresponding to two different irreducible representations, $\Gamma_3$ and $\Gamma_5$. 
$\Gamma_3$ has a single basis vector ($\vec{\psi_1}$) with strictly out-of-plane spins (Fig.~\ref{fig:Neutron}b). 
$\Gamma_5$ has two basis vectors ($\vec{\psi_2}$ and $\vec{\psi_3}$) whose spin components are perpendicular to each other and are strictly pointing within the $ab$-plane (Fig.~\ref{fig:Neutron}c). 
The observation of magnetic Bragg peak at $\vec{Q}$~=~(0,0,L/2) positions (e.g. $\vec{Q}$~=~(0,0,3/2) in Fig.~\ref{fig:Neutron}) rules out the $\Gamma_3$ manifold because neutron scattering is only sensitive to magnetization perpendicular to the momentum transfer $\vec{Q}$. 

The inset of Fig.~\ref{fig:Neutron}a shows that the order parameter (intensity of the $\vec{Q}$~=~(0,0,3/2) peak) approaches the critical point continuously with an exponent $\beta~=~0.45(0.1)$ close to the mean-field value $\beta$ = 0.5.
Thus, the AFM transition in \EZA~is a second-order phase transition, described by a single irreducible representation $\Gamma_5$. 
Due to averaging over symmetrically related magnetic domains, neutron diffraction cannot distinguish between $\psi_2$ and $\psi_3$ basis vectors. 
Also, an accurate estimate of the Eu$^{2+}$ moment size is not possible due to high absorption cross-section of the Eu$^{2+}$ ions. 


\subsection{Magnetoresistance} 
\begin{figure*}
	\includegraphics[width=0.9\textwidth]{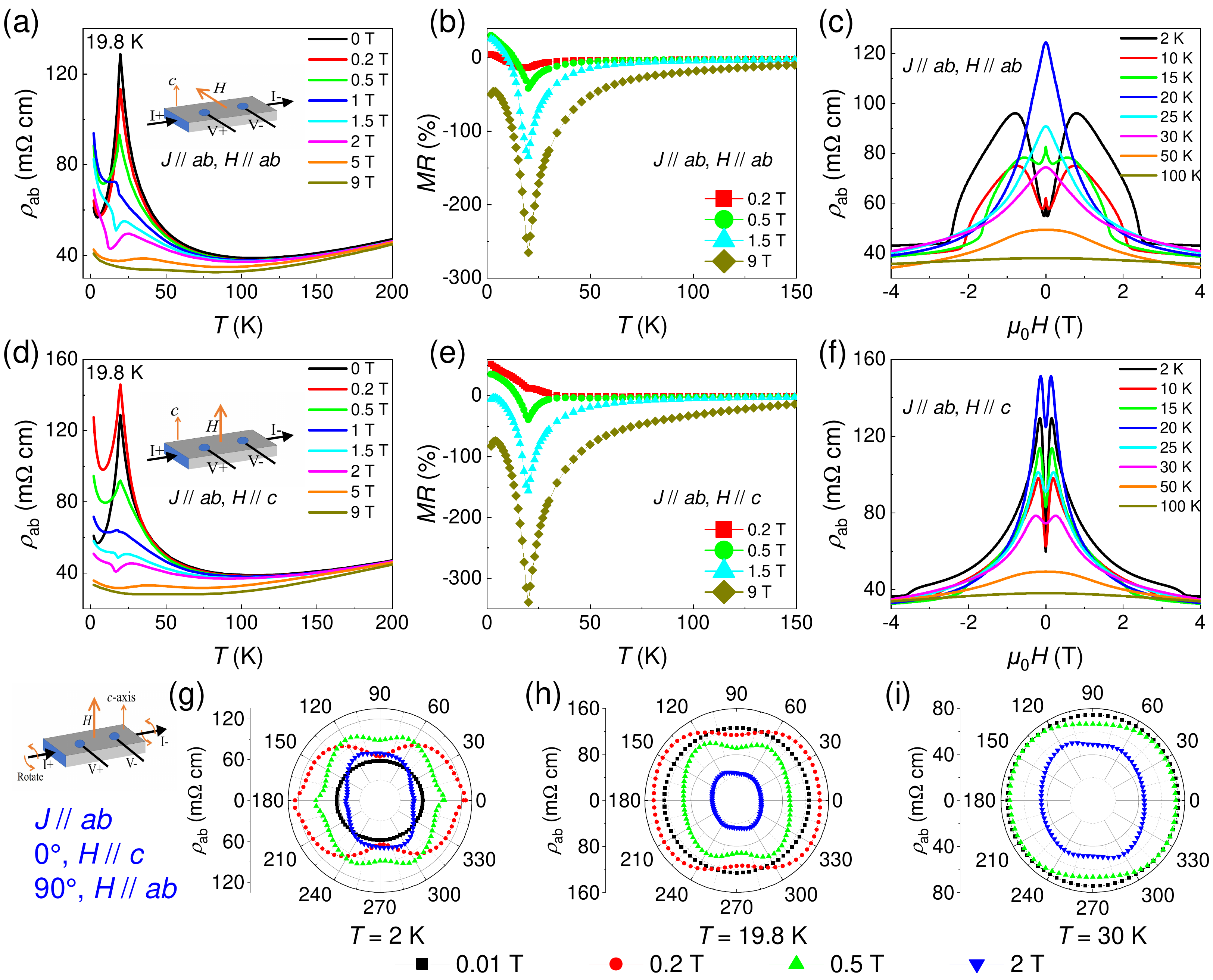}
	\caption{\label{fig:CMR}
		(a) Temperature dependence of in-plane resistivity with fields along $ab$-plane. (b) 
		Magnetoresistance with in-plane fields as a function of temperature. (c) Field dependence of in-plane resistivity at several temperatures. (d)(e)(f) $\rho_\mathrm{ab}(T)$, MR$(T)$ and $\rho_\mathrm{ab}(H)$ with out-of-plane fields, respectively. (g)(h)(i) Angle dependence of in-plane resistivity under several 
		fields at 2 K, 19.8 K and 30 K, respectively.
	}
\end{figure*}

The temperature, field, and angle dependence of in-plane resistivity ($\rho_\mathrm{ab}$) are summarized in Fig.~\ref{fig:CMR}. 
Given the isotropic behavior of \EZA~(Fig.~\ref{fig:ANISOTROPY}), the data for $\rho_\mathrm{c}$ are similar to $\rho_\mathrm{ab}$ as seen in the supplemental Fig.~S4. 
The data collected with in-plane field ($H\parallel ab$) are presented in Figs.~\ref{fig:CMR}a,b,c while Figs.~\ref{fig:CMR}d,e,f show the data with out-of-plane field ($H\parallel c$). 
In either direction, the resistivity of \EZA~shows a negative magnetoresistance (MR) near $T_\textrm{N}$ and a positive MR at small fields.
The two effects are discussed separately below.

\emph{Negative MR near $T_\textrm{N}$.} As discussed earlier, the zero-field resistivity of \EZA\ shows a peak near $T_N$ (Fig.~\ref{fig:ANISOTROPY}e).
This peak is rapidly suppressed when a magnetic field is applied in either $H\|{ab}$ or $H\|c$ direction (Figs.~\ref{fig:CMR}a,d).
We define magnetoresistance as $\mathrm{MR}=100\%\times(\rho(H)-\rho(0))/\rho(H)$ and plot it as a function of temperature at several fields in Figs.~\ref{fig:CMR}b,e. 
The negative MR begins from high temperature ($>100$~K), reaches a maximum of $-340\%$ at $T_\mathrm{N}$ under a 9~T field, and drops to about $-100\%$ below $T_\mathrm{N}$ giving rise to the peak near $T_\textrm{N}$ (Figs.~\ref{fig:CMR}b,e). 
A similar effect has recently been reported in \ECP\ (isostructural to \EZA) but with two differences: 
(i) the negative MR is approximately 20 times larger for in-plane current ($\rho_{ab}$) in \ECP\ and (ii) it shows a ten-fold anisotropy between out-of-plane and in-plane current directions ($\rho_c/\rho_{ab}=10$) unlike the isotropic behavior of \EZA\ (Supplemental Fig.~S4). 
It has been argued that the negative MR in \ECP\ is due to magnetic fluctuations in the region $T_N<T<3T_N$~\cite{AM_EuCd2P2_2021}. 
Within the spin fluctuation mechanism, the smaller magnitude of MR in \EZA\ could be related to the smaller magnetic anisotropy compared to \ECP.
A higher in-plane anisotropy leads to stronger spin fluctuations, higher scattering rate, and larger MR.


\emph{Positive MR at small fields.} In addition to the negative MR, a positive MR is also observed in \EZA\ at low temperatures and under small fields. 
For $H\parallel ab$ and $T=2$ K (the black curve in Fig.~\ref{fig:CMR}c) $\rho_\mathrm{ab}$ increases with increasing field initially, and the positive MR is about 40\% at $H=0.8$ T.
For $\rho_\mathrm{ab}(H_\mathrm{ab})$ curves (Fig.~\ref{fig:CMR}c), the positive MR gradually vanishes as the temperature approaches $T_\textrm{N}=19.6$~K. 
In contrast, for $\rho_\mathrm{ab}(H_\mathrm{c})$ curves, the positive MR survives up to 30 K (Fig.~\ref{fig:CMR}f). 
The positive MR for $\rho_\mathrm{ab}(H_\mathrm{c})$ shows a peak at about 0.15~T at all temperatures before it turns into a negative MR. 
It is worth noting that the magnitude of positive MR at 2~K with $H\parallel c$ is 52\% ($\Delta\rho/\rho(0)=107\%$), i.e. $\rho_\mathrm{ab}$ doubles in a field as small as 0.15~T.

Note that the positive MR depends on the field direction and arises only when the temperature is close to or lower than $T_\mathrm{N}$.
Therefore, it may be related to the close competition between different magnetic structures in \EZA\ shown by our DFT calculations (Fig.~S6).	
The external magnetic field can disturb the balance between competing AFM states (A-type and C-type) and lead to short-range AFM fluctuations that enhance electronic scattering and lead to the positive MR.
A similar mechanism has been proposed in thin film AFM oxides~\cite{positiveMR}.

Figures~\ref{fig:CMR}g,h,i show 360$^\circ$ scans of the resistivity at 2~K, 19.8~K ($T_\mathrm{N}$), and 30~K, respectively. 
The magnitude of the resistivity shows a weak dependence on the field direction near and above $T_\mathrm{N}$ consistent with nearly isotropic magnetization in Fig.~\ref{fig:ANISOTROPY}. 
However, well below $T_\mathrm{N}$ at 2~K, Fig.~\ref{fig:CMR}g shows a 2-fold anisotropy in $\rho_{ab}$ at $H=0.2$~T (red circles).
This observation is consistent with field-induced fluctuations between different magnetic states in \EZA.

\subsection{Heat Capacity} 
\begin{figure*}
	\includegraphics[width=\textwidth]{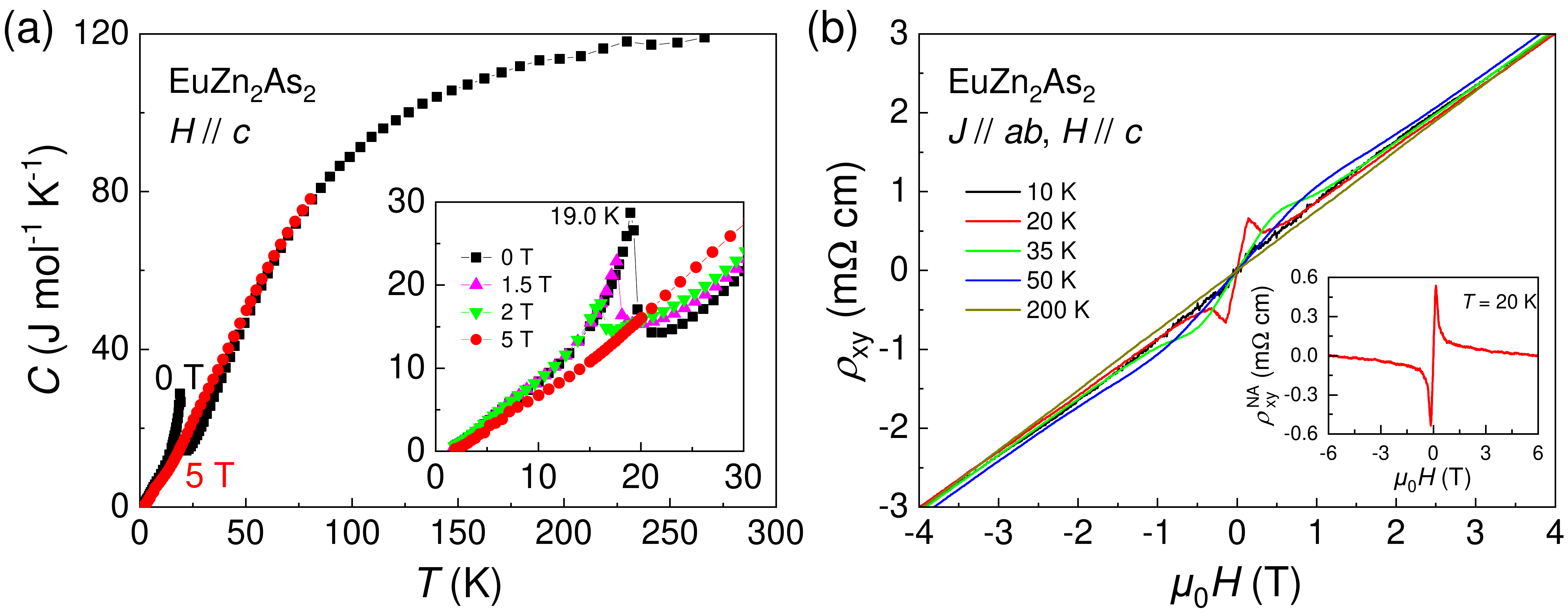}
	\caption{\label{fig:HC}
	(a) Temperature dependence of heat capacity for \EZA\ at several fields. 
	(b) Magnetic field dependence of the Hall resistivity at various temperatures. The fields are along crystallographic $c$ axis while the current is in $ab$ plane.
	}
\end{figure*}
Figure~\ref{fig:HC}a shows the heat capacity of \EZA. 
The peak at 19.0~K is consistent with the AFM transition in $\rho(T)$ and $\chi(T)$ data in Fig.~\ref{fig:ANISOTROPY}. 
As seen in the inset of Fig.~\ref{fig:HC}a, the heat capacity peak is suppressed by increasing the field from zero to 5~T.
The parallel suppression of the magnetic entropy and resistivity with field shows that the negative magnetoresistance in \EZA\ is due to the suppression of spin scattering with increasing field.


\subsection{Non-linear Anomalous Hall Effect} 
The field dependence of the Hall resistivity for \EZA~is shown in Fig.~\ref{fig:HC}b. 
The value of $\rho_{xy}$ increases linearly with the field when the temperature is well above $T_\mathrm{N}$, e.g. at $T=200$~K in Fig.~\ref{fig:HC}b. 
From the positive slope of this ordinary Hall effect (OHE), the carriers are identified as hole-type with the density $n=8.6(2)\times 10^{17}$~cm$^{-1}$.
The field dependence of $\rho_{xy}$ becomes non-linear as the temperature is decreased.
The characteristic peak in the 20~K curve at 0.14~T in Fig.~\ref{fig:HC}b is referred to as the non-linear anomalous Hall effect (NLAHE) because it is not linear in either $B$ or $M$. 
It is diminished rapidly below $T_\mathrm{N}=19.6$~K and vanishes below 10~K.

To analyze the NLAHE, the total Hall resistivity can be expressed as a sum of three contributions, $\rho_{xy}=R_0B+R_SM+\rho_{xy}^\mathrm{NA}$, where $R_0B$ is linear in $B$ and represents the OHE, $R_SM$ is linear in $M$ and represents the conventional anomalous Hall effect (AHE), and $\rho_{xy}^\mathrm{NA}$ is not linear in either $B$ or $M$ and represents the NLAHE~\cite{AHE_EuCd2As2}.
The inset of Fig.~\ref{fig:HC}b shows $\rho_{xy}^\mathrm{NA}$ as a function of field at 20~K after subtracting the OHE and AHE contributions (see also the supplemental Fig.~S5)~\cite{suppmatt}.
At the peak position, the NLAHE constitutes 83\%~of the total Hall resistivity ($\rho_{xy}^\mathrm{NA}/\rho_{xy}=0.83$) in \EZA.
A similar behavior is observed in \ECA, where $\rho_{xy}^\mathrm{NA}/\rho_{xy}=0.97$~\cite{AHE_EuCd2As2,SA_EuCd2As2_2019,EuCd2As2_PRB_2018}.
Both the sheer magnitude of $\rho_{xy}^\mathrm{NA}$ and the ratio $\rho_{xy}^\mathrm{NA}/\rho_{xy}$ in EuX$_2$As$_2$ systems are larger than in materials with a helical magnetic order such as MnSi, MnGe, Fe$_5$Sn$_3$, and Mn$_2$CoAl~\cite{neubauer_topological_2009,kanazawa_large_2011,li_large_2020,ludbrook_nucleation_2017}, and in materials with a large intrinsic (Berry phase) non-linear AHE such as the half-Heusler compounds GdPtBi and DyPtBi~\cite{zhang_field-induced_2020,suzuki_large_2016}. 

\subsection{\label{subsec:dft}Electronic Structure}

\begin{figure*}
	\includegraphics[width=\textwidth]{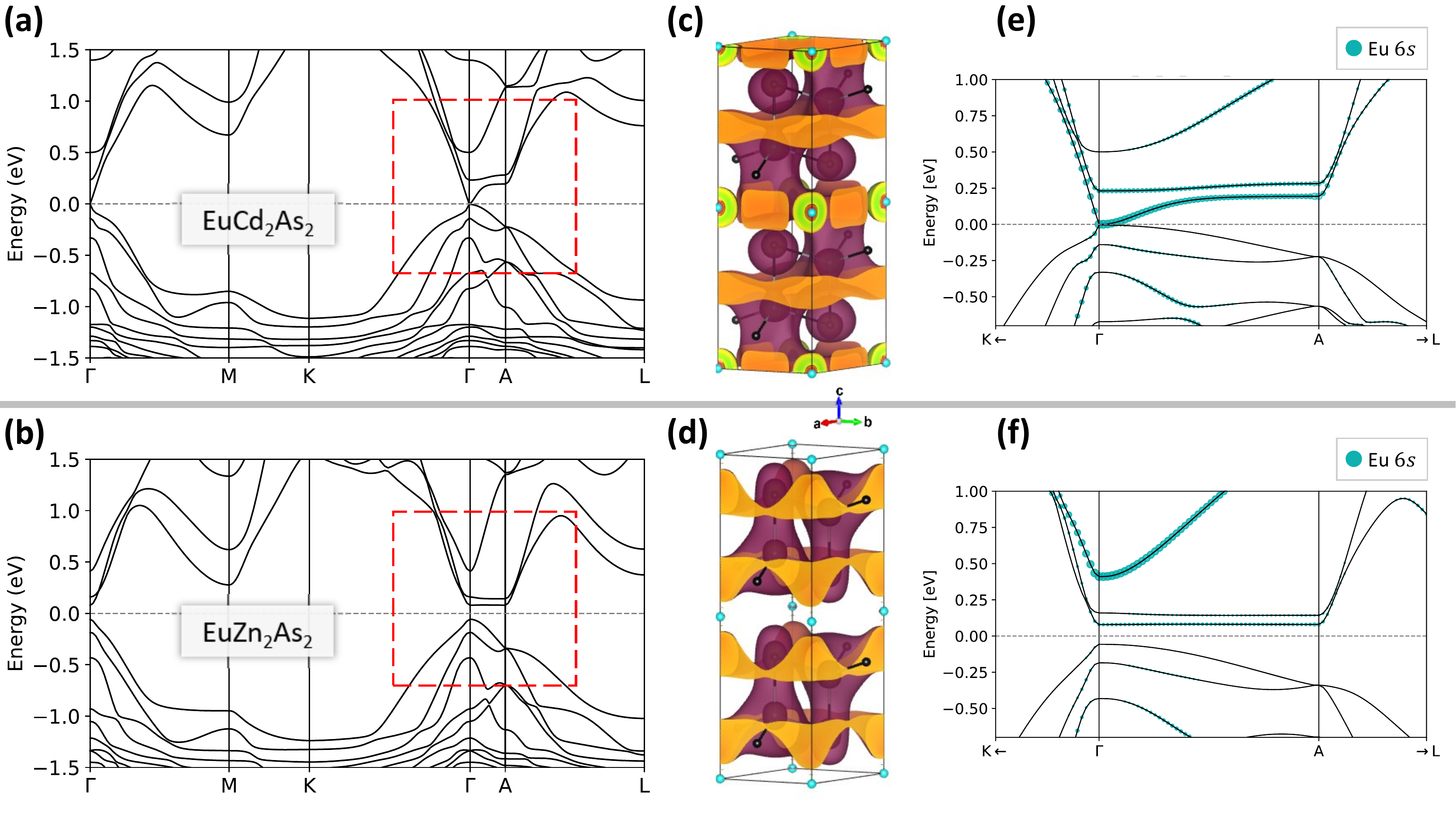}
	\caption{\label{fig:BANDS}
	(a,b) Band structures of \ECA\ and \EZA. The red, dashed box indicates the zoomed-in region in subfigures (e,f).
	(c,d) Charge densities (purple surface) in \ECA\ and \EZA\ for the conduction band at the $\Gamma$-point. Eu, Zn, As atoms are shown as cyan, brown, black balls, respectively.
	(e,f) Band projections onto the Eu-$6s$ orbitals in \ECA\ and \EZA.
	}
\end{figure*}

To investigate the underlying differences between \EZA\ and \ECA, we calculated their band structures assuming type-A AFM order and in-plane spins consistent with experimental findings.
The most significant difference in the electronic structures (Fig.~\ref{fig:BANDS}a,b) is in the band just above the Fermi energy along $\Gamma$-$A$, with \ECA\ showing much greater dispersion and a band touching at $\Gamma$, unlike the gapped and flatter bands in \EZA.
Experimentally \EZA\ is not an insulator, so we expect some amount of disorder and doping to be present in the material.

In order to gain more insight into this interesting region of the band structure, the charge density of the first band above the Fermi level at the $\Gamma$-point are plotted in Fig.~\ref{fig:BANDS}c,d for both \ECA\ and \EZA\ at an isosurface threshold of $2\times 10^{-3}$ electrons. 
Qualitatvely, the frontier orbitals of \ECA\ exhibit an antibonding character within the CdAs$_4$ layer, unlike the bonding character within the ZnAs$_4$ layer in \EZA. 
Another notable difference between the two is the charge density of Eu, being significant in \ECA\ but almost negligible in \EZA.

To better understand the nature of the Eu atom electron density, the projections of Eu-$6s$ orbitals onto the bands are shown in Fig.~\ref{fig:BANDS}e,f for the region indicated by the dashed red squares in Fig.~\ref{fig:BANDS}a,b. 
Even though the overall orbital character of the conduction band is Cd-$5s$ and Zn-$4s$, respectively (see Fig.~S9 and S10), the Eu-$6s$ character provides unique insights. 
The band that is predominantly Eu-$6s$ is different between \ECA\ and \EZA, the former being the first band above the Fermi level, and the latter being the third band. Along with the differences in the Cd(Zn) $4(3)d_{z^2}$ and As $4p_z$ orbital projections onto the band structures (see Fig. S10), the implication is that the third band above the Fermi energy in \EZA\ gets pushed down through the two lower bands to become the first band above the Fermi energy in \ECA.

Considering the dispersion of \ECA\ near $\Gamma$ can help to explain the transport and magnetic anisotropy seen in Section \ref{subsec:Mag}. As can most clearly be seen in the first band above the Fermi level in panel (e),the effective electron mass is small along $K$-$\Gamma$, corresponding to the $ab$-plane direction, compared to the effective mass along $\Gamma$-$A$, corresponding to the $c$-axis direction. This difference in effective electron mass could lead to the anisotropy seen in the experiments.

The origin of the isotropic magnetic and transport measurements in \EZA\ is not as clear cut. Since the first-principles calculations predict \EZA\ to be an insulator, the origin is still an open question although we hypothesize that it is related to a 3-dimensional disorder in the material. Eu can exist in a multivalent state of 2+ and 3+, and given the stoichiometry of these compounds, both valence states of Eu likely exist in each material. In \ECA\, because of the extended nature of the 4$d$-orbitals in Cd, it is likely that the molecular-like orbital formed by Eu and Cd hybridization is more uniform across the material. However, the 3$d$ orbitals of Zn are less extended so cannot smooth the mixture of 2+ and 3+ Eu, leading to 3-dimensional disorder in \EZA\ which would mask any underlying anisotropy. A method to test this hypothesis, which is left to future work, is to measure the $L$-edge of Eu with x-ray absorption spectroscopy (XAS) to determine if the samples of \EZA\ contain a mixture of 2+ and 3+ Eu, leading to 3D disorder.

\section{Conclusions}

The comparative study presented here between \EZA\ and \ECA\ can be summarized as follows.
Both compounds are semimetals with a small concentration of hole carriers ($10^{17-18}$~cm$^{-3}$), hence the large OHE.
Both compounds have A-type AFM order and exhibit a parallel suppression of the magnetic entropy and resistivity with field.
Thus, the negative MR is likely a result of suppressing the spin fluctuations with increasing field.
The main differences between the two compounds is in their respective $T_\textrm{N}$ values and the anisotropy of both magnetic susceptibility and magnetoresistance.
Whereas $\chi_c$ and $\rho_c$ are 4-5 times larger than $\chi_{ab}$ and $\rho_{ab}$ in \ECA\ ($T_\textrm{N}=9.2$~K), they are comparable in \EZA\ ($T_\textrm{N}=19.6$~K).

What underlies this anisotropy difference between \EZA\ and \ECA? 
We can eliminate a structural origin since the two materials have nearly the same $c/a$ ratios, and we can eliminate a magnetic origin since both materials have been shown to have type-A AFM order, leaving the origin to likely be electronic.
Indeed, when comparing the electronic structures from first-principles calculations, there are significant differences between the two materials in their first few bands above the Fermi level. 
The first band above the Fermi level in \ECA\ shows high dispersion and a band touching point at $\Gamma$, has 3D charge density at that point, as is has significant Eu 6$s$ character. 
In contrast, DFT predicts \EZA\ to be gapped with a flat band above the Fermi energy, at $\Gamma$ that band's charge density in separated 2D planes, and the third band above the Fermi energy shows the most significant Eu 6$s$ character. 
The charge and magnetization anisotropy of \ECA\ is likely due to the difference in effective mass of the first band above the Fermi level, which shows very different dispersion in the $ab$-plane versus along the $c$-axis. 
The lack of anisotropy in \EZA\ is still an open question but is likely due to a 3D disorder of Eu valence states which cannot be smoothed by the $d$-orbitals of Zn because they are less extended than those of Cd.

Studying \EZA\ as an analogue to \ECA\ has helped elucidate the effect of tuning between $X=$ Cd, Zn on the temperatures of resistivity peak and NLAHE, and on the anisotropy of magnetization and transport.


\section*{ACKNOWLEDGMENTS}
This material is based upon work supported by the Air Force Office of Scientific Research under award number FA2386-21-1-4059. Any mention of commercial products is intended solely for fully detailing experiments; it does not imply recommendation or endorsement by NIST.


\bibliography{Wang_2Feb2022}

\begin{thebibliography}{38}%
\makeatletter
\providecommand \@ifxundefined [1]{%
 \@ifx{#1\undefined}
}%
\providecommand \@ifnum [1]{%
 \ifnum #1\expandafter \@firstoftwo
 \else \expandafter \@secondoftwo
 \fi
}%
\providecommand \@ifx [1]{%
 \ifx #1\expandafter \@firstoftwo
 \else \expandafter \@secondoftwo
 \fi
}%
\providecommand \natexlab [1]{#1}%
\providecommand \enquote  [1]{``#1''}%
\providecommand \bibnamefont  [1]{#1}%
\providecommand \bibfnamefont [1]{#1}%
\providecommand \citenamefont [1]{#1}%
\providecommand \href@noop [0]{\@secondoftwo}%
\providecommand \href [0]{\begingroup \@sanitize@url \@href}%
\providecommand \@href[1]{\@@startlink{#1}\@@href}%
\providecommand \@@href[1]{\endgroup#1\@@endlink}%
\providecommand \@sanitize@url [0]{\catcode `\\12\catcode `\$12\catcode
  `\&12\catcode `\#12\catcode `\^12\catcode `\_12\catcode `\%12\relax}%
\providecommand \@@startlink[1]{}%
\providecommand \@@endlink[0]{}%
\providecommand \url  [0]{\begingroup\@sanitize@url \@url }%
\providecommand \@url [1]{\endgroup\@href {#1}{\urlprefix }}%
\providecommand \urlprefix  [0]{URL }%
\providecommand \Eprint [0]{\href }%
\providecommand \doibase [0]{http://dx.doi.org/}%
\providecommand \selectlanguage [0]{\@gobble}%
\providecommand \bibinfo  [0]{\@secondoftwo}%
\providecommand \bibfield  [0]{\@secondoftwo}%
\providecommand \translation [1]{[#1]}%
\providecommand \BibitemOpen [0]{}%
\providecommand \bibitemStop [0]{}%
\providecommand \bibitemNoStop [0]{.\EOS\space}%
\providecommand \EOS [0]{\spacefactor3000\relax}%
\providecommand \BibitemShut  [1]{\csname bibitem#1\endcsname}%
\let\auto@bib@innerbib\@empty
\bibitem [{\citenamefont {Wang}\ \emph {et~al.}(2016)\citenamefont {Wang},
  \citenamefont {Wu}, \citenamefont {Shi},\ and\ \citenamefont
  {Wang}}]{EuCd2As2_PRB_2016}%
  \BibitemOpen
  \bibfield  {author} {\bibinfo {author} {\bibfnamefont {H.~P.}\ \bibnamefont
  {Wang}}, \bibinfo {author} {\bibfnamefont {D.~S.}\ \bibnamefont {Wu}},
  \bibinfo {author} {\bibfnamefont {Y.~G.}\ \bibnamefont {Shi}}, \ and\
  \bibinfo {author} {\bibfnamefont {N.~L.}\ \bibnamefont {Wang}},\ }\href
  {\doibase 10.1103/PhysRevB.94.045112} {\bibfield  {journal} {\bibinfo
  {journal} {Phys. Rev. B}\ }\textbf {\bibinfo {volume} {94}},\ \bibinfo
  {pages} {045112} (\bibinfo {year} {2016})}\BibitemShut {NoStop}%
\bibitem [{\citenamefont {Cao}\ \emph {et~al.}(2021)\citenamefont {Cao},
  \citenamefont {Yu}, \citenamefont {Leng}, \citenamefont {Yi}, \citenamefont
  {Yang}, \citenamefont {Liu}, \citenamefont {Kong}, \citenamefont {Li},
  \citenamefont {Dong}, \citenamefont {Shi} \emph {et~al.}}]{AHE_EuCd2As2}%
  \BibitemOpen
  \bibfield  {author} {\bibinfo {author} {\bibfnamefont {X.}~\bibnamefont
  {Cao}}, \bibinfo {author} {\bibfnamefont {J.-X.}\ \bibnamefont {Yu}},
  \bibinfo {author} {\bibfnamefont {P.}~\bibnamefont {Leng}}, \bibinfo {author}
  {\bibfnamefont {C.}~\bibnamefont {Yi}}, \bibinfo {author} {\bibfnamefont
  {Y.}~\bibnamefont {Yang}}, \bibinfo {author} {\bibfnamefont {S.}~\bibnamefont
  {Liu}}, \bibinfo {author} {\bibfnamefont {L.}~\bibnamefont {Kong}}, \bibinfo
  {author} {\bibfnamefont {Z.}~\bibnamefont {Li}}, \bibinfo {author}
  {\bibfnamefont {X.}~\bibnamefont {Dong}}, \bibinfo {author} {\bibfnamefont
  {Y.}~\bibnamefont {Shi}},  \emph {et~al.},\ }\href@noop {} {\bibfield
  {journal} {\bibinfo  {journal} {arXiv:2103.09395}\ } (\bibinfo {year}
  {2021})}\BibitemShut {NoStop}%
\bibitem [{\citenamefont {Ma}\ \emph {et~al.}(2019)\citenamefont {Ma},
  \citenamefont {Nie}, \citenamefont {Yi}, \citenamefont {Jandke},
  \citenamefont {Shang}, \citenamefont {Yao}, \citenamefont {Naamneh},
  \citenamefont {Yan}, \citenamefont {Sun}, \citenamefont {Chikina},
  \citenamefont {Strocov}, \citenamefont {Medarde}, \citenamefont {Song},
  \citenamefont {Xiong}, \citenamefont {Xu}, \citenamefont {Wulfhekel},
  \citenamefont {Mesot}, \citenamefont {Reticcioli}, \citenamefont {Franchini},
  \citenamefont {Mudry}, \citenamefont {M{\"u}ller}, \citenamefont {Shi},
  \citenamefont {Qian}, \citenamefont {Ding},\ and\ \citenamefont
  {Shi}}]{SA_EuCd2As2_2019}%
  \BibitemOpen
  \bibfield  {author} {\bibinfo {author} {\bibfnamefont {J.-Z.}\ \bibnamefont
  {Ma}}, \bibinfo {author} {\bibfnamefont {S.~M.}\ \bibnamefont {Nie}},
  \bibinfo {author} {\bibfnamefont {C.~J.}\ \bibnamefont {Yi}}, \bibinfo
  {author} {\bibfnamefont {J.}~\bibnamefont {Jandke}}, \bibinfo {author}
  {\bibfnamefont {T.}~\bibnamefont {Shang}}, \bibinfo {author} {\bibfnamefont
  {M.~Y.}\ \bibnamefont {Yao}}, \bibinfo {author} {\bibfnamefont
  {M.}~\bibnamefont {Naamneh}}, \bibinfo {author} {\bibfnamefont {L.~Q.}\
  \bibnamefont {Yan}}, \bibinfo {author} {\bibfnamefont {Y.}~\bibnamefont
  {Sun}}, \bibinfo {author} {\bibfnamefont {A.}~\bibnamefont {Chikina}},
  \bibinfo {author} {\bibfnamefont {V.~N.}\ \bibnamefont {Strocov}}, \bibinfo
  {author} {\bibfnamefont {M.}~\bibnamefont {Medarde}}, \bibinfo {author}
  {\bibfnamefont {M.}~\bibnamefont {Song}}, \bibinfo {author} {\bibfnamefont
  {Y.-M.}\ \bibnamefont {Xiong}}, \bibinfo {author} {\bibfnamefont
  {G.}~\bibnamefont {Xu}}, \bibinfo {author} {\bibfnamefont {W.}~\bibnamefont
  {Wulfhekel}}, \bibinfo {author} {\bibfnamefont {J.}~\bibnamefont {Mesot}},
  \bibinfo {author} {\bibfnamefont {M.}~\bibnamefont {Reticcioli}}, \bibinfo
  {author} {\bibfnamefont {C.}~\bibnamefont {Franchini}}, \bibinfo {author}
  {\bibfnamefont {C.}~\bibnamefont {Mudry}}, \bibinfo {author} {\bibfnamefont
  {M.}~\bibnamefont {M{\"u}ller}}, \bibinfo {author} {\bibfnamefont {Y.~G.}\
  \bibnamefont {Shi}}, \bibinfo {author} {\bibfnamefont {T.}~\bibnamefont
  {Qian}}, \bibinfo {author} {\bibfnamefont {H.}~\bibnamefont {Ding}}, \ and\
  \bibinfo {author} {\bibfnamefont {M.}~\bibnamefont {Shi}},\ }\href {\doibase
  10.1126/sciadv.aaw4718} {\bibfield  {journal} {\bibinfo  {journal} {Science
  Advances}\ }\textbf {\bibinfo {volume} {5}} (\bibinfo {year} {2019}),\
  10.1126/sciadv.aaw4718}\BibitemShut {NoStop}%
\bibitem [{\citenamefont {Rahn}\ \emph {et~al.}(2018)\citenamefont {Rahn},
  \citenamefont {Soh}, \citenamefont {Francoual}, \citenamefont {Veiga},
  \citenamefont {Strempfer}, \citenamefont {Mardegan}, \citenamefont {Yan},
  \citenamefont {Guo}, \citenamefont {Shi},\ and\ \citenamefont
  {Boothroyd}}]{EuCd2As2_PRB_2018}%
  \BibitemOpen
  \bibfield  {author} {\bibinfo {author} {\bibfnamefont {M.~C.}\ \bibnamefont
  {Rahn}}, \bibinfo {author} {\bibfnamefont {J.-R.}\ \bibnamefont {Soh}},
  \bibinfo {author} {\bibfnamefont {S.}~\bibnamefont {Francoual}}, \bibinfo
  {author} {\bibfnamefont {L.~S.~I.}\ \bibnamefont {Veiga}}, \bibinfo {author}
  {\bibfnamefont {J.}~\bibnamefont {Strempfer}}, \bibinfo {author}
  {\bibfnamefont {J.}~\bibnamefont {Mardegan}}, \bibinfo {author}
  {\bibfnamefont {D.~Y.}\ \bibnamefont {Yan}}, \bibinfo {author} {\bibfnamefont
  {Y.~F.}\ \bibnamefont {Guo}}, \bibinfo {author} {\bibfnamefont {Y.~G.}\
  \bibnamefont {Shi}}, \ and\ \bibinfo {author} {\bibfnamefont {A.~T.}\
  \bibnamefont {Boothroyd}},\ }\href {\doibase 10.1103/PhysRevB.97.214422}
  {\bibfield  {journal} {\bibinfo  {journal} {Phys. Rev. B}\ }\textbf {\bibinfo
  {volume} {97}},\ \bibinfo {pages} {214422} (\bibinfo {year}
  {2018})}\BibitemShut {NoStop}%
\bibitem [{\citenamefont {Soh}\ \emph {et~al.}(2019)\citenamefont {Soh},
  \citenamefont {de~Juan}, \citenamefont {Vergniory}, \citenamefont
  {Schr\"oter}, \citenamefont {Rahn}, \citenamefont {Yan}, \citenamefont
  {Jiang}, \citenamefont {Bristow}, \citenamefont {Reiss}, \citenamefont
  {Blandy}, \citenamefont {Guo}, \citenamefont {Shi}, \citenamefont {Kim},
  \citenamefont {McCollam}, \citenamefont {Simon}, \citenamefont {Chen},
  \citenamefont {Coldea},\ and\ \citenamefont
  {Boothroyd}}]{PRB_EuCd2As2_2019_JRSoh}%
  \BibitemOpen
  \bibfield  {author} {\bibinfo {author} {\bibfnamefont {J.-R.}\ \bibnamefont
  {Soh}}, \bibinfo {author} {\bibfnamefont {F.}~\bibnamefont {de~Juan}},
  \bibinfo {author} {\bibfnamefont {M.~G.}\ \bibnamefont {Vergniory}}, \bibinfo
  {author} {\bibfnamefont {N.~B.~M.}\ \bibnamefont {Schr\"oter}}, \bibinfo
  {author} {\bibfnamefont {M.~C.}\ \bibnamefont {Rahn}}, \bibinfo {author}
  {\bibfnamefont {D.~Y.}\ \bibnamefont {Yan}}, \bibinfo {author} {\bibfnamefont
  {J.}~\bibnamefont {Jiang}}, \bibinfo {author} {\bibfnamefont
  {M.}~\bibnamefont {Bristow}}, \bibinfo {author} {\bibfnamefont
  {P.}~\bibnamefont {Reiss}}, \bibinfo {author} {\bibfnamefont {J.~N.}\
  \bibnamefont {Blandy}}, \bibinfo {author} {\bibfnamefont {Y.~F.}\
  \bibnamefont {Guo}}, \bibinfo {author} {\bibfnamefont {Y.~G.}\ \bibnamefont
  {Shi}}, \bibinfo {author} {\bibfnamefont {T.~K.}\ \bibnamefont {Kim}},
  \bibinfo {author} {\bibfnamefont {A.}~\bibnamefont {McCollam}}, \bibinfo
  {author} {\bibfnamefont {S.~H.}\ \bibnamefont {Simon}}, \bibinfo {author}
  {\bibfnamefont {Y.}~\bibnamefont {Chen}}, \bibinfo {author} {\bibfnamefont
  {A.~I.}\ \bibnamefont {Coldea}}, \ and\ \bibinfo {author} {\bibfnamefont
  {A.~T.}\ \bibnamefont {Boothroyd}},\ }\href {\doibase
  10.1103/PhysRevB.100.201102} {\bibfield  {journal} {\bibinfo  {journal}
  {Phys. Rev. B}\ }\textbf {\bibinfo {volume} {100}},\ \bibinfo {pages}
  {201102} (\bibinfo {year} {2019})}\BibitemShut {NoStop}%
\bibitem [{\citenamefont {Wang}\ \emph {et~al.}(2019)\citenamefont {Wang},
  \citenamefont {Jo}, \citenamefont {Kuthanazhi}, \citenamefont {Wu},
  \citenamefont {McQueeney}, \citenamefont {Kaminski},\ and\ \citenamefont
  {Canfield}}]{PRB_EuCd2As2_2019_LLWang}%
  \BibitemOpen
  \bibfield  {author} {\bibinfo {author} {\bibfnamefont {L.-L.}\ \bibnamefont
  {Wang}}, \bibinfo {author} {\bibfnamefont {N.~H.}\ \bibnamefont {Jo}},
  \bibinfo {author} {\bibfnamefont {B.}~\bibnamefont {Kuthanazhi}}, \bibinfo
  {author} {\bibfnamefont {Y.}~\bibnamefont {Wu}}, \bibinfo {author}
  {\bibfnamefont {R.~J.}\ \bibnamefont {McQueeney}}, \bibinfo {author}
  {\bibfnamefont {A.}~\bibnamefont {Kaminski}}, \ and\ \bibinfo {author}
  {\bibfnamefont {P.~C.}\ \bibnamefont {Canfield}},\ }\href {\doibase
  10.1103/PhysRevB.99.245147} {\bibfield  {journal} {\bibinfo  {journal} {Phys.
  Rev. B}\ }\textbf {\bibinfo {volume} {99}},\ \bibinfo {pages} {245147}
  (\bibinfo {year} {2019})}\BibitemShut {NoStop}%
\bibitem [{\citenamefont {Ma}\ \emph {et~al.}(2020)\citenamefont {Ma},
  \citenamefont {Wang}, \citenamefont {Nie}, \citenamefont {Yi}, \citenamefont
  {Xu}, \citenamefont {Li}, \citenamefont {Jandke}, \citenamefont {Wulfhekel},
  \citenamefont {Huang}, \citenamefont {West}, \citenamefont {Richard},
  \citenamefont {Chikina}, \citenamefont {Strocov}, \citenamefont {Mesot},
  \citenamefont {Weng}, \citenamefont {Zhang}, \citenamefont {Shi},
  \citenamefont {Qian}, \citenamefont {Shi},\ and\ \citenamefont
  {Ding}}]{Ma_EuCd2As2_2020}%
  \BibitemOpen
  \bibfield  {author} {\bibinfo {author} {\bibfnamefont {J.}~\bibnamefont
  {Ma}}, \bibinfo {author} {\bibfnamefont {H.}~\bibnamefont {Wang}}, \bibinfo
  {author} {\bibfnamefont {S.}~\bibnamefont {Nie}}, \bibinfo {author}
  {\bibfnamefont {C.}~\bibnamefont {Yi}}, \bibinfo {author} {\bibfnamefont
  {Y.}~\bibnamefont {Xu}}, \bibinfo {author} {\bibfnamefont {H.}~\bibnamefont
  {Li}}, \bibinfo {author} {\bibfnamefont {J.}~\bibnamefont {Jandke}}, \bibinfo
  {author} {\bibfnamefont {W.}~\bibnamefont {Wulfhekel}}, \bibinfo {author}
  {\bibfnamefont {Y.}~\bibnamefont {Huang}}, \bibinfo {author} {\bibfnamefont
  {D.}~\bibnamefont {West}}, \bibinfo {author} {\bibfnamefont {P.}~\bibnamefont
  {Richard}}, \bibinfo {author} {\bibfnamefont {A.}~\bibnamefont {Chikina}},
  \bibinfo {author} {\bibfnamefont {V.~N.}\ \bibnamefont {Strocov}}, \bibinfo
  {author} {\bibfnamefont {J.}~\bibnamefont {Mesot}}, \bibinfo {author}
  {\bibfnamefont {H.}~\bibnamefont {Weng}}, \bibinfo {author} {\bibfnamefont
  {S.}~\bibnamefont {Zhang}}, \bibinfo {author} {\bibfnamefont
  {Y.}~\bibnamefont {Shi}}, \bibinfo {author} {\bibfnamefont {T.}~\bibnamefont
  {Qian}}, \bibinfo {author} {\bibfnamefont {M.}~\bibnamefont {Shi}}, \ and\
  \bibinfo {author} {\bibfnamefont {H.}~\bibnamefont {Ding}},\ }\href {\doibase
  https://doi.org/10.1002/adma.201907565} {\bibfield  {journal} {\bibinfo
  {journal} {Advanced Materials}\ }\textbf {\bibinfo {volume} {32}},\ \bibinfo
  {pages} {1907565} (\bibinfo {year} {2020})}\BibitemShut {NoStop}%
\bibitem [{\citenamefont {Soh}\ \emph {et~al.}(2020)\citenamefont {Soh},
  \citenamefont {Schierle}, \citenamefont {Yan}, \citenamefont {Su},
  \citenamefont {Prabhakaran}, \citenamefont {Weschke}, \citenamefont {Guo},
  \citenamefont {Shi},\ and\ \citenamefont {Boothroyd}}]{soh_resonant_2020}%
  \BibitemOpen
  \bibfield  {author} {\bibinfo {author} {\bibfnamefont {J.-R.}\ \bibnamefont
  {Soh}}, \bibinfo {author} {\bibfnamefont {E.}~\bibnamefont {Schierle}},
  \bibinfo {author} {\bibfnamefont {D.~Y.}\ \bibnamefont {Yan}}, \bibinfo
  {author} {\bibfnamefont {H.}~\bibnamefont {Su}}, \bibinfo {author}
  {\bibfnamefont {D.}~\bibnamefont {Prabhakaran}}, \bibinfo {author}
  {\bibfnamefont {E.}~\bibnamefont {Weschke}}, \bibinfo {author} {\bibfnamefont
  {Y.~F.}\ \bibnamefont {Guo}}, \bibinfo {author} {\bibfnamefont {Y.~G.}\
  \bibnamefont {Shi}}, \ and\ \bibinfo {author} {\bibfnamefont {A.~T.}\
  \bibnamefont {Boothroyd}},\ }\href {\doibase 10.1103/PhysRevB.102.014408}
  {\bibfield  {journal} {\bibinfo  {journal} {Physical Review B}\ }\textbf
  {\bibinfo {volume} {102}},\ \bibinfo {pages} {014408} (\bibinfo {year}
  {2020})}\BibitemShut {NoStop}%
\bibitem [{\citenamefont {Schiffer}\ \emph {et~al.}(1995)\citenamefont
  {Schiffer}, \citenamefont {Ramirez}, \citenamefont {Bao},\ and\ \citenamefont
  {Cheong}}]{LaMnO}%
  \BibitemOpen
  \bibfield  {author} {\bibinfo {author} {\bibfnamefont {P.}~\bibnamefont
  {Schiffer}}, \bibinfo {author} {\bibfnamefont {A.~P.}\ \bibnamefont
  {Ramirez}}, \bibinfo {author} {\bibfnamefont {W.}~\bibnamefont {Bao}}, \ and\
  \bibinfo {author} {\bibfnamefont {S.-W.}\ \bibnamefont {Cheong}},\ }\href
  {\doibase 10.1103/PhysRevLett.75.3336} {\bibfield  {journal} {\bibinfo
  {journal} {Phys. Rev. Lett.}\ }\textbf {\bibinfo {volume} {75}},\ \bibinfo
  {pages} {3336} (\bibinfo {year} {1995})}\BibitemShut {NoStop}%
\bibitem [{\citenamefont {Subramanian}\ \emph {et~al.}(1996)\citenamefont
  {Subramanian}, \citenamefont {Toby}, \citenamefont {Ramirez}, \citenamefont
  {Marshall}, \citenamefont {Sleight},\ and\ \citenamefont {Kwei}}]{Tl2Mn2O7}%
  \BibitemOpen
  \bibfield  {author} {\bibinfo {author} {\bibfnamefont {M.~A.}\ \bibnamefont
  {Subramanian}}, \bibinfo {author} {\bibfnamefont {B.~H.}\ \bibnamefont
  {Toby}}, \bibinfo {author} {\bibfnamefont {A.~P.}\ \bibnamefont {Ramirez}},
  \bibinfo {author} {\bibfnamefont {W.~J.}\ \bibnamefont {Marshall}}, \bibinfo
  {author} {\bibfnamefont {A.~W.}\ \bibnamefont {Sleight}}, \ and\ \bibinfo
  {author} {\bibfnamefont {G.~H.}\ \bibnamefont {Kwei}},\ }\href {\doibase
  10.1126/science.273.5271.81} {\bibfield  {journal} {\bibinfo  {journal}
  {Science}\ }\textbf {\bibinfo {volume} {273}},\ \bibinfo {pages} {81}
  (\bibinfo {year} {1996})}\BibitemShut {NoStop}%
\bibitem [{\citenamefont {Ramirez}\ \emph {et~al.}(1997)\citenamefont
  {Ramirez}, \citenamefont {Cava},\ and\ \citenamefont {Krajewski}}]{FeCr2S4}%
  \BibitemOpen
  \bibfield  {author} {\bibinfo {author} {\bibfnamefont {A.}~\bibnamefont
  {Ramirez}}, \bibinfo {author} {\bibfnamefont {R.~J.}\ \bibnamefont {Cava}}, \
  and\ \bibinfo {author} {\bibfnamefont {J.}~\bibnamefont {Krajewski}},\
  }\href@noop {} {\bibfield  {journal} {\bibinfo  {journal} {Nature}\ }\textbf
  {\bibinfo {volume} {386}},\ \bibinfo {pages} {156} (\bibinfo {year}
  {1997})}\BibitemShut {NoStop}%
\bibitem [{\citenamefont {Chan}\ \emph {et~al.}(1997)\citenamefont {Chan},
  \citenamefont {Kauzlarich}, \citenamefont {Klavins}, \citenamefont
  {Shelton},\ and\ \citenamefont {Webb}}]{Eu14MnSb11}%
  \BibitemOpen
  \bibfield  {author} {\bibinfo {author} {\bibfnamefont {J.~Y.}\ \bibnamefont
  {Chan}}, \bibinfo {author} {\bibfnamefont {S.~M.}\ \bibnamefont
  {Kauzlarich}}, \bibinfo {author} {\bibfnamefont {P.}~\bibnamefont {Klavins}},
  \bibinfo {author} {\bibfnamefont {R.~N.}\ \bibnamefont {Shelton}}, \ and\
  \bibinfo {author} {\bibfnamefont {D.~J.}\ \bibnamefont {Webb}},\ }\href@noop
  {} {\bibfield  {journal} {\bibinfo  {journal} {Chemistry of materials}\
  }\textbf {\bibinfo {volume} {9}},\ \bibinfo {pages} {3132} (\bibinfo {year}
  {1997})}\BibitemShut {NoStop}%
\bibitem [{\citenamefont {Fisher}\ \emph {et~al.}(1999)\citenamefont {Fisher},
  \citenamefont {Wiener}, \citenamefont {Bud'ko}, \citenamefont {Canfield},
  \citenamefont {Chan},\ and\ \citenamefont {Kauzlarich}}]{Yb14MnSb11}%
  \BibitemOpen
  \bibfield  {author} {\bibinfo {author} {\bibfnamefont {I.~R.}\ \bibnamefont
  {Fisher}}, \bibinfo {author} {\bibfnamefont {T.~A.}\ \bibnamefont {Wiener}},
  \bibinfo {author} {\bibfnamefont {S.~L.}\ \bibnamefont {Bud'ko}}, \bibinfo
  {author} {\bibfnamefont {P.~C.}\ \bibnamefont {Canfield}}, \bibinfo {author}
  {\bibfnamefont {J.~Y.}\ \bibnamefont {Chan}}, \ and\ \bibinfo {author}
  {\bibfnamefont {S.~M.}\ \bibnamefont {Kauzlarich}},\ }\href {\doibase
  10.1103/PhysRevB.59.13829} {\bibfield  {journal} {\bibinfo  {journal} {Phys.
  Rev. B}\ }\textbf {\bibinfo {volume} {59}},\ \bibinfo {pages} {13829}
  (\bibinfo {year} {1999})}\BibitemShut {NoStop}%
\bibitem [{\citenamefont {Jiang}\ and\ \citenamefont
  {Kauzlarich}(2006)}]{EuIn2P2}%
  \BibitemOpen
  \bibfield  {author} {\bibinfo {author} {\bibfnamefont {J.}~\bibnamefont
  {Jiang}}\ and\ \bibinfo {author} {\bibfnamefont {S.~M.}\ \bibnamefont
  {Kauzlarich}},\ }\href@noop {} {\bibfield  {journal} {\bibinfo  {journal}
  {Chemistry of materials}\ }\textbf {\bibinfo {volume} {18}},\ \bibinfo
  {pages} {435} (\bibinfo {year} {2006})}\BibitemShut {NoStop}%
\bibitem [{\citenamefont {Goforth}\ \emph {et~al.}(2008)\citenamefont
  {Goforth}, \citenamefont {Klavins}, \citenamefont {Fettinger},\ and\
  \citenamefont {Kauzlarich}}]{EuIn2As2}%
  \BibitemOpen
  \bibfield  {author} {\bibinfo {author} {\bibfnamefont {A.~M.}\ \bibnamefont
  {Goforth}}, \bibinfo {author} {\bibfnamefont {P.}~\bibnamefont {Klavins}},
  \bibinfo {author} {\bibfnamefont {J.~C.}\ \bibnamefont {Fettinger}}, \ and\
  \bibinfo {author} {\bibfnamefont {S.~M.}\ \bibnamefont {Kauzlarich}},\
  }\href@noop {} {\bibfield  {journal} {\bibinfo  {journal} {Inorganic
  chemistry}\ }\textbf {\bibinfo {volume} {47}},\ \bibinfo {pages} {11048}
  (\bibinfo {year} {2008})}\BibitemShut {NoStop}%
\bibitem [{\citenamefont
  {Rodr\'{i}guez-Carvajal}(1993)}]{rodriguez-carvajal_recent_1993}%
  \BibitemOpen
  \bibfield  {author} {\bibinfo {author} {\bibfnamefont {J.}~\bibnamefont
  {Rodr\'{i}guez-Carvajal}},\ }\href {\doibase 10.1016/0921-4526(93)90108-I}
  {\bibfield  {journal} {\bibinfo  {journal} {Physica B: Condensed Matter}\
  }\textbf {\bibinfo {volume} {192}},\ \bibinfo {pages} {55} (\bibinfo {year}
  {1993})}\BibitemShut {NoStop}%
\bibitem [{\citenamefont {Wills}(2000)}]{SARAh}%
  \BibitemOpen
  \bibfield  {author} {\bibinfo {author} {\bibfnamefont {A.}~\bibnamefont
  {Wills}},\ }\href {\doibase https://doi.org/10.1016/S0921-4526(99)01722-6}
  {\bibfield  {journal} {\bibinfo  {journal} {Physica B: Condensed Matter}\
  }\textbf {\bibinfo {volume} {276-278}},\ \bibinfo {pages} {680} (\bibinfo
  {year} {2000})}\BibitemShut {NoStop}%
\bibitem [{\citenamefont {Kresse}\ and\ \citenamefont
  {Joubert}(1999)}]{VASP-PAW_Kresse_Joubert_PRB1999}%
  \BibitemOpen
  \bibfield  {author} {\bibinfo {author} {\bibfnamefont {G.}~\bibnamefont
  {Kresse}}\ and\ \bibinfo {author} {\bibfnamefont {D.}~\bibnamefont
  {Joubert}},\ }\href {\doibase 10.1103/PhysRevB.59.1758} {\bibfield  {journal}
  {\bibinfo  {journal} {Phys. Rev. B}\ }\textbf {\bibinfo {volume} {59}},\
  \bibinfo {pages} {1758} (\bibinfo {year} {1999})}\BibitemShut {NoStop}%
\bibitem [{\citenamefont {Perdew}\ \emph {et~al.}(1996)\citenamefont {Perdew},
  \citenamefont {Burke},\ and\ \citenamefont {Ernzerhof}}]{PBE}%
  \BibitemOpen
  \bibfield  {author} {\bibinfo {author} {\bibfnamefont {J.~P.}\ \bibnamefont
  {Perdew}}, \bibinfo {author} {\bibfnamefont {K.}~\bibnamefont {Burke}}, \
  and\ \bibinfo {author} {\bibfnamefont {M.}~\bibnamefont {Ernzerhof}},\ }\href
  {\doibase 10.1103/PhysRevLett.77.3865} {\bibfield  {journal} {\bibinfo
  {journal} {Phys. Rev. Lett.}\ }\textbf {\bibinfo {volume} {77}},\ \bibinfo
  {pages} {3865} (\bibinfo {year} {1996})}\BibitemShut {NoStop}%
\bibitem [{\citenamefont {Kresse}\ and\ \citenamefont
  {Furthm\"uller}(1996{\natexlab{a}})}]{VASP_Kresse_Furthmuller_CompMatSci1996}%
  \BibitemOpen
  \bibfield  {author} {\bibinfo {author} {\bibfnamefont {G.}~\bibnamefont
  {Kresse}}\ and\ \bibinfo {author} {\bibfnamefont {J.}~\bibnamefont
  {Furthm\"uller}},\ }\href {\doibase
  https://doi.org/10.1016/0927-0256(96)00008-0} {\bibfield  {journal} {\bibinfo
   {journal} {Computational Materials Science}\ }\textbf {\bibinfo {volume}
  {6}},\ \bibinfo {pages} {15} (\bibinfo {year}
  {1996}{\natexlab{a}})}\BibitemShut {NoStop}%
\bibitem [{\citenamefont {Kresse}\ and\ \citenamefont
  {Furthm\"uller}(1996{\natexlab{b}})}]{VASP_Kresse_Furthmuller_PRB1996}%
  \BibitemOpen
  \bibfield  {author} {\bibinfo {author} {\bibfnamefont {G.}~\bibnamefont
  {Kresse}}\ and\ \bibinfo {author} {\bibfnamefont {J.}~\bibnamefont
  {Furthm\"uller}},\ }\href {\doibase 10.1103/PhysRevB.54.11169} {\bibfield
  {journal} {\bibinfo  {journal} {Phys. Rev. B}\ }\textbf {\bibinfo {volume}
  {54}},\ \bibinfo {pages} {11169} (\bibinfo {year}
  {1996}{\natexlab{b}})}\BibitemShut {NoStop}%
\bibitem [{\citenamefont {Kresse}\ and\ \citenamefont
  {Hafner}(1993)}]{VASP_Kresse_Hafner_PRB1993}%
  \BibitemOpen
  \bibfield  {author} {\bibinfo {author} {\bibfnamefont {G.}~\bibnamefont
  {Kresse}}\ and\ \bibinfo {author} {\bibfnamefont {J.}~\bibnamefont
  {Hafner}},\ }\href {\doibase 10.1103/PhysRevB.47.558} {\bibfield  {journal}
  {\bibinfo  {journal} {Phys. Rev. B}\ }\textbf {\bibinfo {volume} {47}},\
  \bibinfo {pages} {558} (\bibinfo {year} {1993})}\BibitemShut {NoStop}%
\bibitem [{\citenamefont {Kresse}\ and\ \citenamefont
  {Hafner}(1994)}]{VASP_Kresse_Hafner_PRB1994}%
  \BibitemOpen
  \bibfield  {author} {\bibinfo {author} {\bibfnamefont {G.}~\bibnamefont
  {Kresse}}\ and\ \bibinfo {author} {\bibfnamefont {J.}~\bibnamefont
  {Hafner}},\ }\href {\doibase 10.1103/PhysRevB.49.14251} {\bibfield  {journal}
  {\bibinfo  {journal} {Phys. Rev. B}\ }\textbf {\bibinfo {volume} {49}},\
  \bibinfo {pages} {14251} (\bibinfo {year} {1994})}\BibitemShut {NoStop}%
\bibitem [{\citenamefont {Xu}\ \emph {et~al.}(2019)\citenamefont {Xu},
  \citenamefont {Song}, \citenamefont {Wang}, \citenamefont {Weng},\ and\
  \citenamefont {Dai}}]{EuIn2As2_axionInsulator}%
  \BibitemOpen
  \bibfield  {author} {\bibinfo {author} {\bibfnamefont {Y.}~\bibnamefont
  {Xu}}, \bibinfo {author} {\bibfnamefont {Z.}~\bibnamefont {Song}}, \bibinfo
  {author} {\bibfnamefont {Z.}~\bibnamefont {Wang}}, \bibinfo {author}
  {\bibfnamefont {H.}~\bibnamefont {Weng}}, \ and\ \bibinfo {author}
  {\bibfnamefont {X.}~\bibnamefont {Dai}},\ }\href {\doibase
  10.1103/PhysRevLett.122.256402} {\bibfield  {journal} {\bibinfo  {journal}
  {Phys. Rev. Lett.}\ }\textbf {\bibinfo {volume} {122}},\ \bibinfo {pages}
  {256402} (\bibinfo {year} {2019})}\BibitemShut {NoStop}%
\bibitem [{\citenamefont {Dudarev}\ \emph {et~al.}(1998)\citenamefont
  {Dudarev}, \citenamefont {Botton}, \citenamefont {Savrasov}, \citenamefont
  {Humphreys},\ and\ \citenamefont {Sutton}}]{LDAUTYPE2}%
  \BibitemOpen
  \bibfield  {author} {\bibinfo {author} {\bibfnamefont {S.~L.}\ \bibnamefont
  {Dudarev}}, \bibinfo {author} {\bibfnamefont {G.~A.}\ \bibnamefont {Botton}},
  \bibinfo {author} {\bibfnamefont {S.~Y.}\ \bibnamefont {Savrasov}}, \bibinfo
  {author} {\bibfnamefont {C.~J.}\ \bibnamefont {Humphreys}}, \ and\ \bibinfo
  {author} {\bibfnamefont {A.~P.}\ \bibnamefont {Sutton}},\ }\href {\doibase
  10.1103/PhysRevB.57.1505} {\bibfield  {journal} {\bibinfo  {journal} {Phys.
  Rev. B}\ }\textbf {\bibinfo {volume} {57}},\ \bibinfo {pages} {1505}
  (\bibinfo {year} {1998})}\BibitemShut {NoStop}%
\bibitem [{sup()}]{suppmatt}%
  \BibitemOpen
  \href {https://journals.aps.org} {}\bibinfo {note} {See the Supplemental
  Material for details}\BibitemShut {NoStop}%
\bibitem [{\citenamefont {Dimmock}\ and\ \citenamefont
  {Freeman}(1964)}]{PRL_fdcouple}%
  \BibitemOpen
  \bibfield  {author} {\bibinfo {author} {\bibfnamefont {J.~O.}\ \bibnamefont
  {Dimmock}}\ and\ \bibinfo {author} {\bibfnamefont {A.~J.}\ \bibnamefont
  {Freeman}},\ }\href {\doibase 10.1103/PhysRevLett.13.750} {\bibfield
  {journal} {\bibinfo  {journal} {Phys. Rev. Lett.}\ }\textbf {\bibinfo
  {volume} {13}},\ \bibinfo {pages} {750} (\bibinfo {year} {1964})}\BibitemShut
  {NoStop}%
\bibitem [{\citenamefont {Li}\ \emph {et~al.}(1995)\citenamefont {Li},
  \citenamefont {Pearson}, \citenamefont {Bader}, \citenamefont {McIlroy},
  \citenamefont {Waldfried},\ and\ \citenamefont {Dowben}}]{PRB_fdcouple}%
  \BibitemOpen
  \bibfield  {author} {\bibinfo {author} {\bibfnamefont {D.}~\bibnamefont
  {Li}}, \bibinfo {author} {\bibfnamefont {J.}~\bibnamefont {Pearson}},
  \bibinfo {author} {\bibfnamefont {S.~D.}\ \bibnamefont {Bader}}, \bibinfo
  {author} {\bibfnamefont {D.~N.}\ \bibnamefont {McIlroy}}, \bibinfo {author}
  {\bibfnamefont {C.}~\bibnamefont {Waldfried}}, \ and\ \bibinfo {author}
  {\bibfnamefont {P.~A.}\ \bibnamefont {Dowben}},\ }\href {\doibase
  10.1103/PhysRevB.51.13895} {\bibfield  {journal} {\bibinfo  {journal} {Phys.
  Rev. B}\ }\textbf {\bibinfo {volume} {51}},\ \bibinfo {pages} {13895}
  (\bibinfo {year} {1995})}\BibitemShut {NoStop}%
\bibitem [{\citenamefont {Ioffe}\ and\ \citenamefont
  {Regel}(1960)}]{ioffe_non-crystalline_1960}%
  \BibitemOpen
  \bibfield  {author} {\bibinfo {author} {\bibfnamefont {A.~F.}\ \bibnamefont
  {Ioffe}}\ and\ \bibinfo {author} {\bibfnamefont {A.~R.}\ \bibnamefont
  {Regel}},\ }\href {https://www.elibrary.ru/item.asp?id=21765390} {\ ,\
  \bibinfo {pages} {237} (\bibinfo {year} {1960})}\BibitemShut {NoStop}%
\bibitem [{\citenamefont {Emery}\ and\ \citenamefont
  {Kivelson}(1995)}]{PRL_badmetal}%
  \BibitemOpen
  \bibfield  {author} {\bibinfo {author} {\bibfnamefont {V.~J.}\ \bibnamefont
  {Emery}}\ and\ \bibinfo {author} {\bibfnamefont {S.~A.}\ \bibnamefont
  {Kivelson}},\ }\href {\doibase 10.1103/PhysRevLett.74.3253} {\bibfield
  {journal} {\bibinfo  {journal} {Phys. Rev. Lett.}\ }\textbf {\bibinfo
  {volume} {74}},\ \bibinfo {pages} {3253} (\bibinfo {year}
  {1995})}\BibitemShut {NoStop}%
\bibitem [{\citenamefont {Wang}\ \emph {et~al.}()\citenamefont {Wang},
  \citenamefont {Rogers}, \citenamefont {Yao}, \citenamefont {Nichols},
  \citenamefont {Atay}, \citenamefont {Xu}, \citenamefont {Franklin},
  \citenamefont {Sochnikov}, \citenamefont {Ryan}, \citenamefont {Haskel},\
  and\ \citenamefont {Tafti}}]{AM_EuCd2P2_2021}%
  \BibitemOpen
  \bibfield  {author} {\bibinfo {author} {\bibfnamefont {Z.-C.}\ \bibnamefont
  {Wang}}, \bibinfo {author} {\bibfnamefont {J.~D.}\ \bibnamefont {Rogers}},
  \bibinfo {author} {\bibfnamefont {X.}~\bibnamefont {Yao}}, \bibinfo {author}
  {\bibfnamefont {R.}~\bibnamefont {Nichols}}, \bibinfo {author} {\bibfnamefont
  {K.}~\bibnamefont {Atay}}, \bibinfo {author} {\bibfnamefont {B.}~\bibnamefont
  {Xu}}, \bibinfo {author} {\bibfnamefont {J.}~\bibnamefont {Franklin}},
  \bibinfo {author} {\bibfnamefont {I.}~\bibnamefont {Sochnikov}}, \bibinfo
  {author} {\bibfnamefont {P.~J.}\ \bibnamefont {Ryan}}, \bibinfo {author}
  {\bibfnamefont {D.}~\bibnamefont {Haskel}}, \ and\ \bibinfo {author}
  {\bibfnamefont {F.}~\bibnamefont {Tafti}},\ }\href {\doibase
  https://doi.org/10.1002/adma.202005755} {\bibfield  {journal} {\bibinfo
  {journal} {Advanced Materials}\ }\textbf {\bibinfo {volume} {n/a}},\ \bibinfo
  {pages} {2005755}}\BibitemShut {NoStop}%
\bibitem [{\citenamefont {Zhang}\ \emph {et~al.}(2017)\citenamefont {Zhang},
  \citenamefont {Ji}, \citenamefont {Xu}, \citenamefont {Geng}, \citenamefont
  {Zhou}, \citenamefont {Gu}, \citenamefont {Yao},\ and\ \citenamefont
  {Zhang}}]{positiveMR}%
  \BibitemOpen
  \bibfield  {author} {\bibinfo {author} {\bibfnamefont {J.}~\bibnamefont
  {Zhang}}, \bibinfo {author} {\bibfnamefont {W.-J.}\ \bibnamefont {Ji}},
  \bibinfo {author} {\bibfnamefont {J.}~\bibnamefont {Xu}}, \bibinfo {author}
  {\bibfnamefont {X.-Y.}\ \bibnamefont {Geng}}, \bibinfo {author}
  {\bibfnamefont {J.}~\bibnamefont {Zhou}}, \bibinfo {author} {\bibfnamefont
  {Z.-B.}\ \bibnamefont {Gu}}, \bibinfo {author} {\bibfnamefont {S.-H.}\
  \bibnamefont {Yao}}, \ and\ \bibinfo {author} {\bibfnamefont {S.-T.}\
  \bibnamefont {Zhang}},\ }\href {\doibase 10.1126/sciadv.1701473} {\bibfield
  {journal} {\bibinfo  {journal} {Science Advances}\ }\textbf {\bibinfo
  {volume} {3}} (\bibinfo {year} {2017}),\ 10.1126/sciadv.1701473},\ \Eprint
  {http://arxiv.org/abs/https://advances.sciencemag.org/content/3/11/e1701473.full.pdf}
  {https://advances.sciencemag.org/content/3/11/e1701473.full.pdf} \BibitemShut
  {NoStop}%
\bibitem [{\citenamefont {Neubauer}\ \emph {et~al.}(2009)\citenamefont
  {Neubauer}, \citenamefont {Pfleiderer}, \citenamefont {Binz}, \citenamefont
  {Rosch}, \citenamefont {Ritz}, \citenamefont {Niklowitz},\ and\ \citenamefont
  {Böni}}]{neubauer_topological_2009}%
  \BibitemOpen
  \bibfield  {author} {\bibinfo {author} {\bibfnamefont {A.}~\bibnamefont
  {Neubauer}}, \bibinfo {author} {\bibfnamefont {C.}~\bibnamefont
  {Pfleiderer}}, \bibinfo {author} {\bibfnamefont {B.}~\bibnamefont {Binz}},
  \bibinfo {author} {\bibfnamefont {A.}~\bibnamefont {Rosch}}, \bibinfo
  {author} {\bibfnamefont {R.}~\bibnamefont {Ritz}}, \bibinfo {author}
  {\bibfnamefont {P.~G.}\ \bibnamefont {Niklowitz}}, \ and\ \bibinfo {author}
  {\bibfnamefont {P.}~\bibnamefont {Böni}},\ }\href {\doibase
  10.1103/PhysRevLett.102.186602} {\bibfield  {journal} {\bibinfo  {journal}
  {Physical Review Letters}\ }\textbf {\bibinfo {volume} {102}},\ \bibinfo
  {pages} {186602} (\bibinfo {year} {2009})}\BibitemShut {NoStop}%
\bibitem [{\citenamefont {Kanazawa}\ \emph {et~al.}(2011)\citenamefont
  {Kanazawa}, \citenamefont {Onose}, \citenamefont {Arima}, \citenamefont
  {Okuyama}, \citenamefont {Ohoyama}, \citenamefont {Wakimoto}, \citenamefont
  {Kakurai}, \citenamefont {Ishiwata},\ and\ \citenamefont
  {Tokura}}]{kanazawa_large_2011}%
  \BibitemOpen
  \bibfield  {author} {\bibinfo {author} {\bibfnamefont {N.}~\bibnamefont
  {Kanazawa}}, \bibinfo {author} {\bibfnamefont {Y.}~\bibnamefont {Onose}},
  \bibinfo {author} {\bibfnamefont {T.}~\bibnamefont {Arima}}, \bibinfo
  {author} {\bibfnamefont {D.}~\bibnamefont {Okuyama}}, \bibinfo {author}
  {\bibfnamefont {K.}~\bibnamefont {Ohoyama}}, \bibinfo {author} {\bibfnamefont
  {S.}~\bibnamefont {Wakimoto}}, \bibinfo {author} {\bibfnamefont
  {K.}~\bibnamefont {Kakurai}}, \bibinfo {author} {\bibfnamefont
  {S.}~\bibnamefont {Ishiwata}}, \ and\ \bibinfo {author} {\bibfnamefont
  {Y.}~\bibnamefont {Tokura}},\ }\href {\doibase
  10.1103/PhysRevLett.106.156603} {\bibfield  {journal} {\bibinfo  {journal}
  {Physical Review Letters}\ }\textbf {\bibinfo {volume} {106}},\ \bibinfo
  {pages} {156603} (\bibinfo {year} {2011})}\BibitemShut {NoStop}%
\bibitem [{\citenamefont {Li}\ \emph {et~al.}(2020)\citenamefont {Li},
  \citenamefont {Ding}, \citenamefont {Chen}, \citenamefont {Li}, \citenamefont
  {Liu}, \citenamefont {Xi}, \citenamefont {Wu},\ and\ \citenamefont
  {Wang}}]{li_large_2020}%
  \BibitemOpen
  \bibfield  {author} {\bibinfo {author} {\bibfnamefont {H.}~\bibnamefont
  {Li}}, \bibinfo {author} {\bibfnamefont {B.}~\bibnamefont {Ding}}, \bibinfo
  {author} {\bibfnamefont {J.}~\bibnamefont {Chen}}, \bibinfo {author}
  {\bibfnamefont {Z.}~\bibnamefont {Li}}, \bibinfo {author} {\bibfnamefont
  {E.}~\bibnamefont {Liu}}, \bibinfo {author} {\bibfnamefont {X.}~\bibnamefont
  {Xi}}, \bibinfo {author} {\bibfnamefont {G.}~\bibnamefont {Wu}}, \ and\
  \bibinfo {author} {\bibfnamefont {W.}~\bibnamefont {Wang}},\ }\href {\doibase
  10.1063/5.0005493} {\bibfield  {journal} {\bibinfo  {journal} {Applied
  Physics Letters}\ }\textbf {\bibinfo {volume} {116}},\ \bibinfo {pages}
  {182405} (\bibinfo {year} {2020})}\BibitemShut {NoStop}%
\bibitem [{\citenamefont {Ludbrook}\ \emph {et~al.}(2017)\citenamefont
  {Ludbrook}, \citenamefont {Dubuis}, \citenamefont {Puichaud}, \citenamefont
  {Ruck},\ and\ \citenamefont {Granville}}]{ludbrook_nucleation_2017}%
  \BibitemOpen
  \bibfield  {author} {\bibinfo {author} {\bibfnamefont {B.~M.}\ \bibnamefont
  {Ludbrook}}, \bibinfo {author} {\bibfnamefont {G.}~\bibnamefont {Dubuis}},
  \bibinfo {author} {\bibfnamefont {A.-H.}\ \bibnamefont {Puichaud}}, \bibinfo
  {author} {\bibfnamefont {B.~J.}\ \bibnamefont {Ruck}}, \ and\ \bibinfo
  {author} {\bibfnamefont {S.}~\bibnamefont {Granville}},\ }\href {\doibase
  10.1038/s41598-017-13211-8} {\bibfield  {journal} {\bibinfo  {journal}
  {Scientific Reports}\ }\textbf {\bibinfo {volume} {7}},\ \bibinfo {pages}
  {13620} (\bibinfo {year} {2017})}\BibitemShut {NoStop}%
\bibitem [{\citenamefont {Zhang}\ \emph {et~al.}(2020)\citenamefont {Zhang},
  \citenamefont {Zhu}, \citenamefont {Qiu}, \citenamefont {Tian}, \citenamefont
  {Cao}, \citenamefont {Mao},\ and\ \citenamefont
  {Ke}}]{zhang_field-induced_2020}%
  \BibitemOpen
  \bibfield  {author} {\bibinfo {author} {\bibfnamefont {H.}~\bibnamefont
  {Zhang}}, \bibinfo {author} {\bibfnamefont {Y.~L.}\ \bibnamefont {Zhu}},
  \bibinfo {author} {\bibfnamefont {Y.}~\bibnamefont {Qiu}}, \bibinfo {author}
  {\bibfnamefont {W.}~\bibnamefont {Tian}}, \bibinfo {author} {\bibfnamefont
  {H.~B.}\ \bibnamefont {Cao}}, \bibinfo {author} {\bibfnamefont {Z.~Q.}\
  \bibnamefont {Mao}}, \ and\ \bibinfo {author} {\bibfnamefont
  {X.}~\bibnamefont {Ke}},\ }\href {\doibase 10.1103/PhysRevB.102.094424}
  {\bibfield  {journal} {\bibinfo  {journal} {Physical Review B}\ }\textbf
  {\bibinfo {volume} {102}},\ \bibinfo {pages} {094424} (\bibinfo {year}
  {2020})}\BibitemShut {NoStop}%
\bibitem [{\citenamefont {Suzuki}\ \emph {et~al.}(2016)\citenamefont {Suzuki},
  \citenamefont {Chisnell}, \citenamefont {Devarakonda}, \citenamefont {Liu},
  \citenamefont {Feng}, \citenamefont {Xiao}, \citenamefont {Lynn},\ and\
  \citenamefont {Checkelsky}}]{suzuki_large_2016}%
  \BibitemOpen
  \bibfield  {author} {\bibinfo {author} {\bibfnamefont {T.}~\bibnamefont
  {Suzuki}}, \bibinfo {author} {\bibfnamefont {R.}~\bibnamefont {Chisnell}},
  \bibinfo {author} {\bibfnamefont {A.}~\bibnamefont {Devarakonda}}, \bibinfo
  {author} {\bibfnamefont {Y.-T.}\ \bibnamefont {Liu}}, \bibinfo {author}
  {\bibfnamefont {W.}~\bibnamefont {Feng}}, \bibinfo {author} {\bibfnamefont
  {D.}~\bibnamefont {Xiao}}, \bibinfo {author} {\bibfnamefont {J.~W.}\
  \bibnamefont {Lynn}}, \ and\ \bibinfo {author} {\bibfnamefont {J.~G.}\
  \bibnamefont {Checkelsky}},\ }\href {\doibase 10.1038/nphys3831} {\bibfield
  {journal} {\bibinfo  {journal} {Nature Physics}\ }\textbf {\bibinfo {volume}
  {12}},\ \bibinfo {pages} {1119} (\bibinfo {year} {2016})}\BibitemShut
  {NoStop}%
\end{thebibliography}%


\begin{thebibliography}{3}%
\makeatletter
\providecommand \@ifxundefined [1]{%
 \@ifx{#1\undefined}
}%
\providecommand \@ifnum [1]{%
 \ifnum #1\expandafter \@firstoftwo
 \else \expandafter \@secondoftwo
 \fi
}%
\providecommand \@ifx [1]{%
 \ifx #1\expandafter \@firstoftwo
 \else \expandafter \@secondoftwo
 \fi
}%
\providecommand \natexlab [1]{#1}%
\providecommand \enquote  [1]{``#1''}%
\providecommand \bibnamefont  [1]{#1}%
\providecommand \bibfnamefont [1]{#1}%
\providecommand \citenamefont [1]{#1}%
\providecommand \href@noop [0]{\@secondoftwo}%
\providecommand \href [0]{\begingroup \@sanitize@url \@href}%
\providecommand \@href[1]{\@@startlink{#1}\@@href}%
\providecommand \@@href[1]{\endgroup#1\@@endlink}%
\providecommand \@sanitize@url [0]{\catcode `\\12\catcode `\$12\catcode
  `\&12\catcode `\#12\catcode `\^12\catcode `\_12\catcode `\%12\relax}%
\providecommand \@@startlink[1]{}%
\providecommand \@@endlink[0]{}%
\providecommand \url  [0]{\begingroup\@sanitize@url \@url }%
\providecommand \@url [1]{\endgroup\@href {#1}{\urlprefix }}%
\providecommand \urlprefix  [0]{URL }%
\providecommand \Eprint [0]{\href }%
\providecommand \doibase [0]{http://dx.doi.org/}%
\providecommand \selectlanguage [0]{\@gobble}%
\providecommand \bibinfo  [0]{\@secondoftwo}%
\providecommand \bibfield  [0]{\@secondoftwo}%
\providecommand \translation [1]{[#1]}%
\providecommand \BibitemOpen [0]{}%
\providecommand \bibitemStop [0]{}%
\providecommand \bibitemNoStop [0]{.\EOS\space}%
\providecommand \EOS [0]{\spacefactor3000\relax}%
\providecommand \BibitemShut  [1]{\csname bibitem#1\endcsname}%
\let\auto@bib@innerbib\@empty
\bibitem [{\citenamefont {Wang}\ \emph {et~al.}(2017)\citenamefont {Wang},
  \citenamefont {Li}, \citenamefont {Wang}, \citenamefont {Li}, \citenamefont
  {Zhang}, \citenamefont {Zhou}, \citenamefont {Chen},\ and\ \citenamefont
  {Pei}}]{EuZn2Sb2_Uvalue}%
  \BibitemOpen
  \bibfield  {author} {\bibinfo {author} {\bibfnamefont {X.}~\bibnamefont
  {Wang}}, \bibinfo {author} {\bibfnamefont {W.}~\bibnamefont {Li}}, \bibinfo
  {author} {\bibfnamefont {C.}~\bibnamefont {Wang}}, \bibinfo {author}
  {\bibfnamefont {J.}~\bibnamefont {Li}}, \bibinfo {author} {\bibfnamefont
  {X.}~\bibnamefont {Zhang}}, \bibinfo {author} {\bibfnamefont
  {B.}~\bibnamefont {Zhou}}, \bibinfo {author} {\bibfnamefont {Y.}~\bibnamefont
  {Chen}}, \ and\ \bibinfo {author} {\bibfnamefont {Y.}~\bibnamefont {Pei}},\
  }\href {\doibase 10.1039/C7TA08869H} {\bibfield  {journal} {\bibinfo
  {journal} {J. Mater. Chem. A}\ }\textbf {\bibinfo {volume} {5}},\ \bibinfo
  {pages} {24185} (\bibinfo {year} {2017})}\BibitemShut {NoStop}%
\bibitem [{\citenamefont {Sun}\ \emph {et~al.}(2016)\citenamefont {Sun},
  \citenamefont {Remsing}, \citenamefont {Zhang}, \citenamefont {Sun},
  \citenamefont {Ruzsinszky}, \citenamefont {Peng}, \citenamefont {Yang},
  \citenamefont {Paul}, \citenamefont {Waghmare}, \citenamefont {Wu},
  \citenamefont {Klein},\ and\ \citenamefont {Perdew}}]{SCAN_performance}%
  \BibitemOpen
  \bibfield  {author} {\bibinfo {author} {\bibfnamefont {J.}~\bibnamefont
  {Sun}}, \bibinfo {author} {\bibfnamefont {R.~C.}\ \bibnamefont {Remsing}},
  \bibinfo {author} {\bibfnamefont {Y.}~\bibnamefont {Zhang}}, \bibinfo
  {author} {\bibfnamefont {Z.}~\bibnamefont {Sun}}, \bibinfo {author}
  {\bibfnamefont {A.}~\bibnamefont {Ruzsinszky}}, \bibinfo {author}
  {\bibfnamefont {H.}~\bibnamefont {Peng}}, \bibinfo {author} {\bibfnamefont
  {Z.}~\bibnamefont {Yang}}, \bibinfo {author} {\bibfnamefont {A.}~\bibnamefont
  {Paul}}, \bibinfo {author} {\bibfnamefont {U.}~\bibnamefont {Waghmare}},
  \bibinfo {author} {\bibfnamefont {X.}~\bibnamefont {Wu}}, \bibinfo {author}
  {\bibfnamefont {M.~L.}\ \bibnamefont {Klein}}, \ and\ \bibinfo {author}
  {\bibfnamefont {J.~P.}\ \bibnamefont {Perdew}},\ }\href {\doibase
  10.1038/nchem.2535} {\bibfield  {journal} {\bibinfo  {journal} {Nature
  Chemistry}\ }\textbf {\bibinfo {volume} {8}},\ \bibinfo {pages} {831}
  (\bibinfo {year} {2016})}\BibitemShut {NoStop}%
\bibitem [{\citenamefont {Sun}\ \emph {et~al.}(2015)\citenamefont {Sun},
  \citenamefont {Ruzsinszky},\ and\ \citenamefont {Perdew}}]{SCAN_original}%
  \BibitemOpen
  \bibfield  {author} {\bibinfo {author} {\bibfnamefont {J.}~\bibnamefont
  {Sun}}, \bibinfo {author} {\bibfnamefont {A.}~\bibnamefont {Ruzsinszky}}, \
  and\ \bibinfo {author} {\bibfnamefont {J.~P.}\ \bibnamefont {Perdew}},\
  }\href {\doibase 10.1103/PhysRevLett.115.036402} {\bibfield  {journal}
  {\bibinfo  {journal} {Phys. Rev. Lett.}\ }\textbf {\bibinfo {volume} {115}},\
  \bibinfo {pages} {036402} (\bibinfo {year} {2015})}\BibitemShut {NoStop}%
\end{thebibliography}%

\end{document}



\title{Supplemental Materials: Anisotropy of the magnetic and transport properties in EuZn$_2$As$_2$}

\author{Zhi-Cheng~Wang}
\thanks{The authors contributed equally to this work.}
\affiliation{Department of Physics, Boston College, Chestnut Hill, MA 02467, USA}

\author{Emily Been}
\thanks{The authors contributed equally to this work.}
\affiliation{Stanford Institute for Materials and Energy Sciences, SLAC National Accelerator Laboratory, 2575 Sand Hill Road, Menlo Park, CA 94025.}
\affiliation{Department of Physics, Stanford University, Stanford, CA 94305.}

\author{Jonathan~Gaudet}
\affiliation{NIST Center for Neutron Research, National Institute of Standards and Technology, Gaithersburg, Maryland 20899, USA}
\affiliation{Department of Materials Science and Eng., University of Maryland, College Park, MD 20742-2115}

\author{Gadeer~Matook~A.~Alqasseri} 
\affiliation{Department of Physics and Astrophysics, Howard University, Washington DC, 20059; 2IBM-Howard Quantum Center, Howard University, Washington DC, 20059.}

\author{Kyle~Fruhling}
\author{Xiaohan~Yao}
\affiliation{Department of Physics, Boston College, Chestnut Hill, MA 02467, USA}

\author{Uwe~Stuhr}
\affiliation{Paul Scherrer Institut, 5232 Villigen, Switzerland}

\author{Qinqing Zhu} 
\author{Zhi Ren}
\affiliation{School of Science, Westlake University, 18 Shilongshan Road, Hangzhou 310064, PR China.}
\affiliation{Institute of Natural Sciences, Westlake Institute for Advanced Study, 18 Shilongshan Road, Hangzhou 310064, PR China.}

\author{Yi Cui}
\affiliation{Department of Materials Science and Engineering, Stanford University, Stanford, CA 94305.}

\author{Chunjing Jia} 
\affiliation{Stanford Institute for Materials and Energy Sciences, SLAC National Accelerator Laboratory, 2575 Sand Hill Road, Menlo Park, CA 94025.}

\author{Brian Moritz}
\affiliation{Stanford Institute for Materials and Energy Sciences, SLAC National Accelerator Laboratory, 2575 Sand Hill Road, Menlo Park, CA 94025.}

\author{Sugata~Chowdhury} 
\affiliation{Department of Physics and Astrophysics, Howard University, Washington DC, 20059; 2IBM-Howard Quantum Center, Howard University, Washington DC, 20059.}
\affiliation{IBM-Howard Quantum Center, Howard University, Washington DC, 20059.}

\author{Thomas Devereaux}
\affiliation{Stanford Institute for Materials and Energy Sciences, SLAC National Accelerator Laboratory, 2575 Sand Hill Road, Menlo Park, CA 94025.}
\affiliation{Department of Materials Science and Engineering, Stanford University, Stanford, CA 94305.}

\author{Fazel~Tafti}
\email{fazel.tafti@bc.edu}
\affiliation{Department of Physics, Boston College, Chestnut Hill, MA 02467, USA}

\date{\today}

\maketitle

\section{XRD}\label{sec:appendix-xrd}
\begin{figure}
	\centering
	\includegraphics[width=0.6\columnwidth]{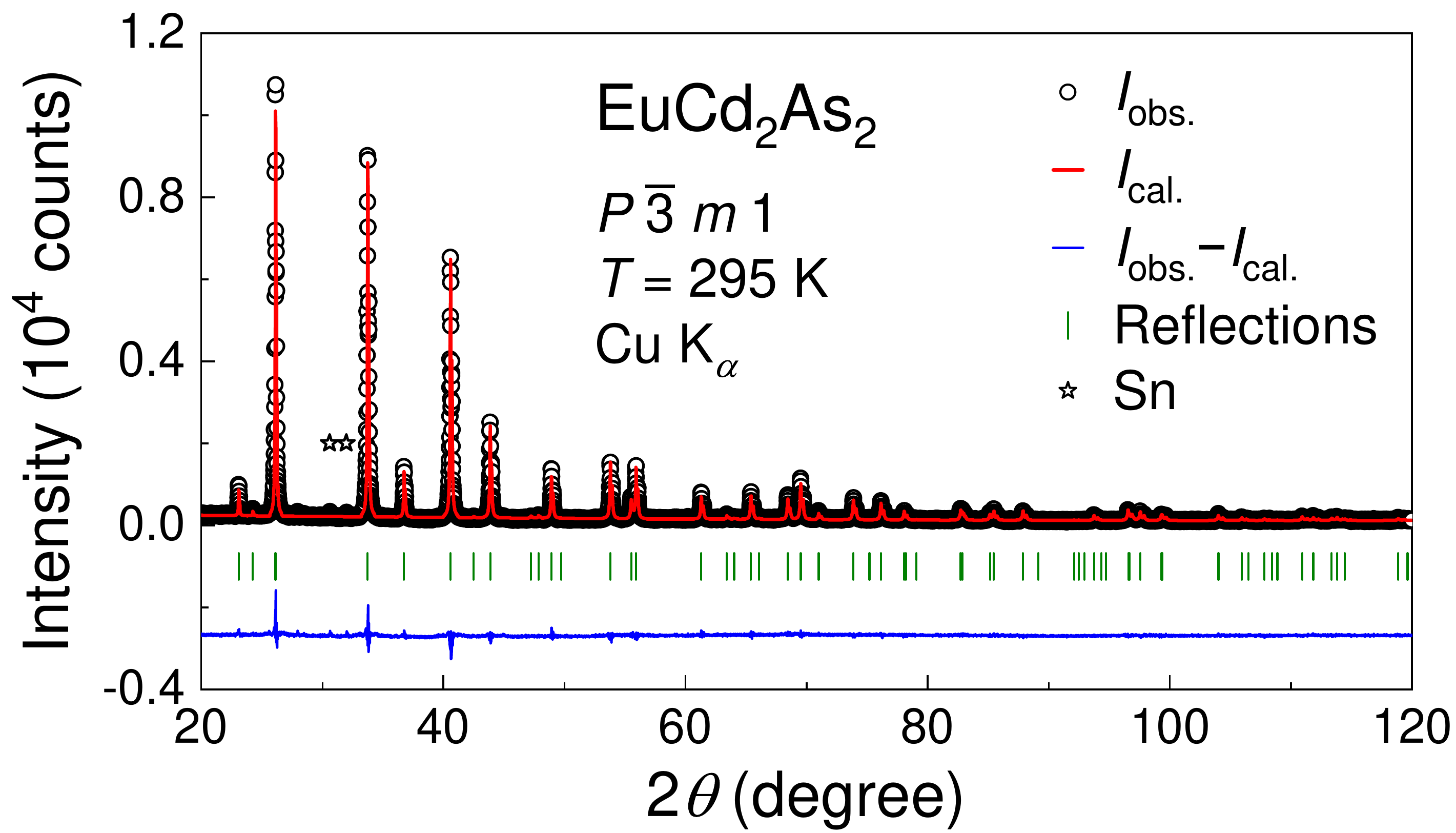}
	\caption{Powder X-ray diffraction data and the Rietveld analysis on \ECA\ samples. The asterisks mark two peaks from the tin flux used to grow the crystals.}
	\label{fig:XRD}
\end{figure}
The information in Table 1 of the main text comes from the Rietveld analysis of powder X-ray diffraction data from our \ECA\ samples presented in Fig.~\ref{fig:XRD}. 
Since \ECA\ has been studied before, we did not present the data in the main text.

\section{Hysteresis}\label{sec:appendix-hysteresis}
\begin{figure*}
	\centering
	\includegraphics[width=\columnwidth]{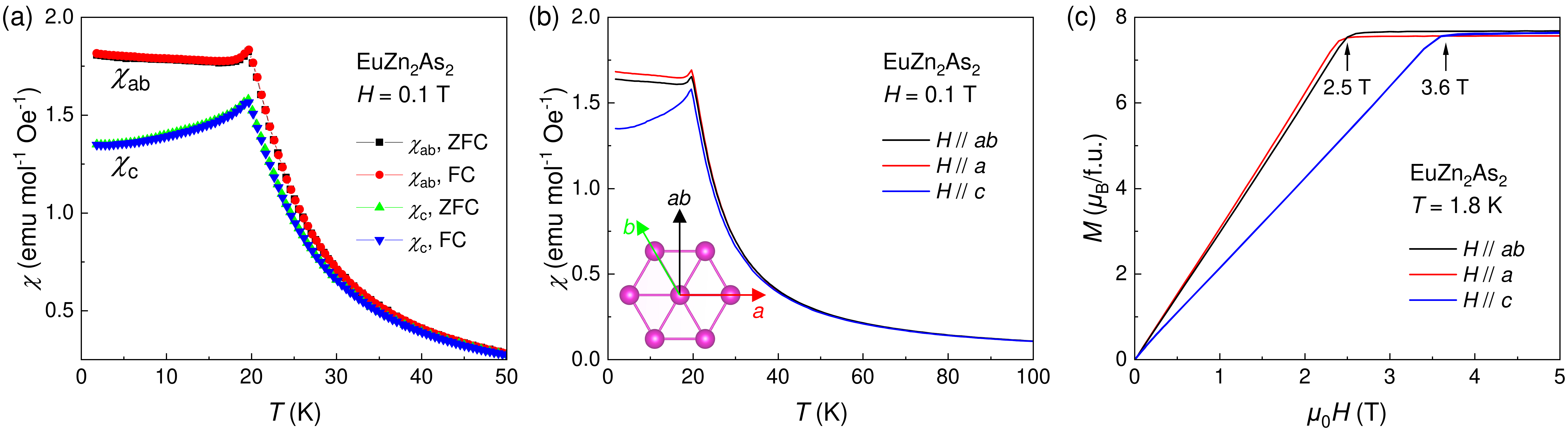}
	\caption{(a) The ZFC and FC curves for anisotropic magnetic susceptibility with in-plane and out-of-plane field directions.
	(b) Temperature dependence of anisotropic magnetic susceptibility with in-plane fields ($H\| ab$, $H\| a$) and out-of-plane field ($H\| c$).
    (c) Field dependence of magnetization with in-plane fields ($H\| ab$, $H\| a$) and out-of-plane field ($H\| c$) at 1.8 K. }
	\label{fig:ZFCFC}
\end{figure*}
It was mentioned in the main text that there is no splitting between the ZFC and FC susceptibility curves in \EZA, but only the ZFC data were presented for clarity.
Here, we show both ZFC and FC data in Fig.~\ref{fig:ZFCFC}a under both in-plane and out-of-plane magnetic fields.
The FC and ZFC curves for each field direction are on top of each other.

In the main text, we showed the in-plane susceptibility and magnetization data ($\chi_{ab}$ and $M(H\| ab)$ without specifying the exact in-plane direction. 
This is because there is no in-plane anisotropy in the magnetic properties of \EZA\ as seen in Figs.~\ref{fig:ZFCFC}b,c.
The temperature dependence of $\chi_{a}$ and $\chi_{ab}$ are identical in Fig.~\ref{fig:ZFCFC}b, and so are the field dependences of $M(H\|a)$ and $M(H\|ab)$ in Fig.~\ref{fig:ZFCFC}c.

\section{Curie-Weiss Analysis}\label{sec:cw-analysis}
\begin{figure}
	\centering
	\includegraphics[width=0.7\columnwidth]{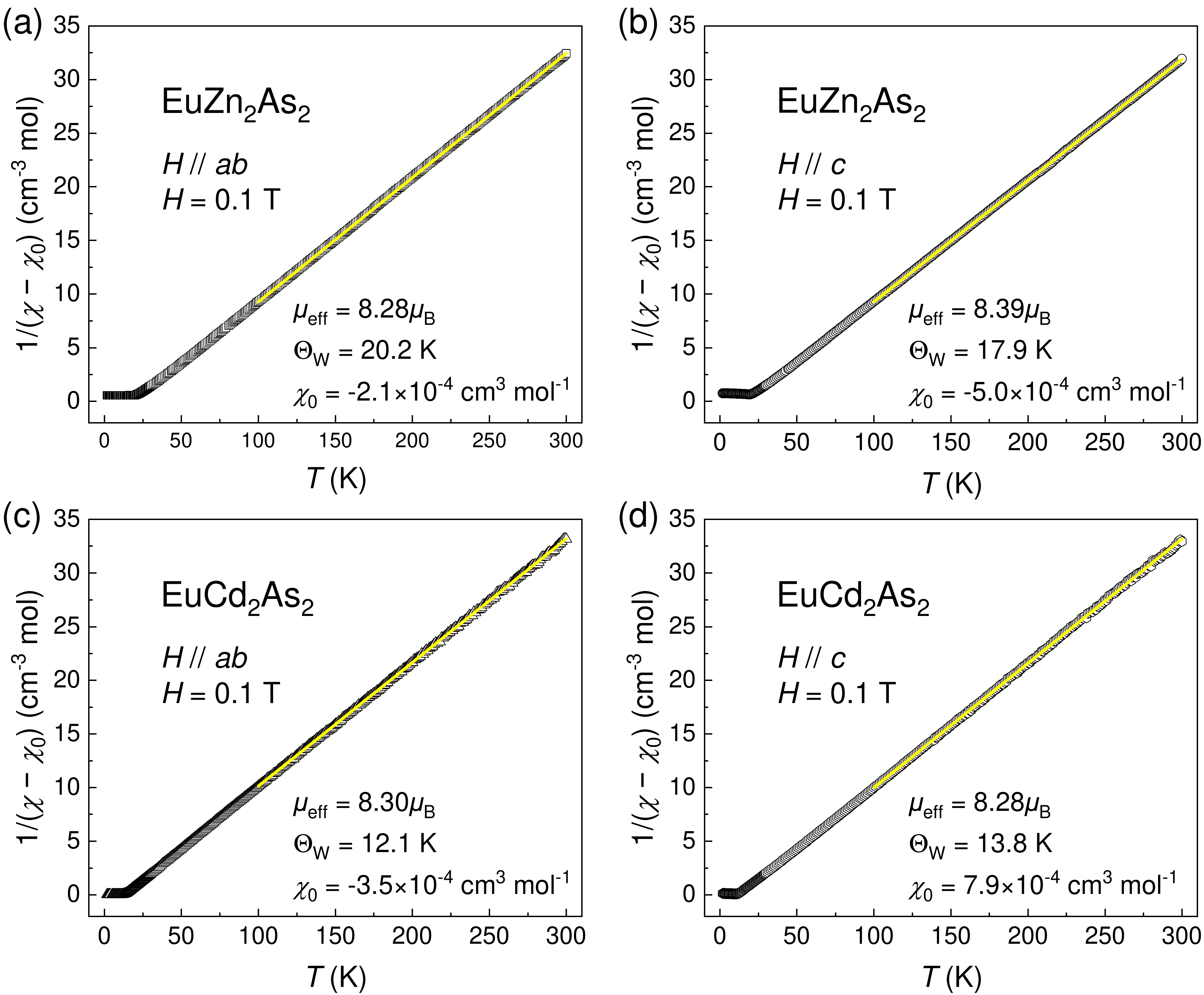}
	\caption{The Curie-Weiss analysis in both \ECA\ and \EZA\ with different field directions.}
	\label{fig:CWfit}
\end{figure}
The Curie-Weiss analysis is presented for both \ECA\ and \EZA\ with both $H\|c$ and $H\|ab$.
The analysis shows comparable values of the magnetic moment and Weiss temperature in either direction.

\section{Out-of-plane resistivity under fields}
\begin{figure*}[h]
	\includegraphics[width=0.9\textwidth]{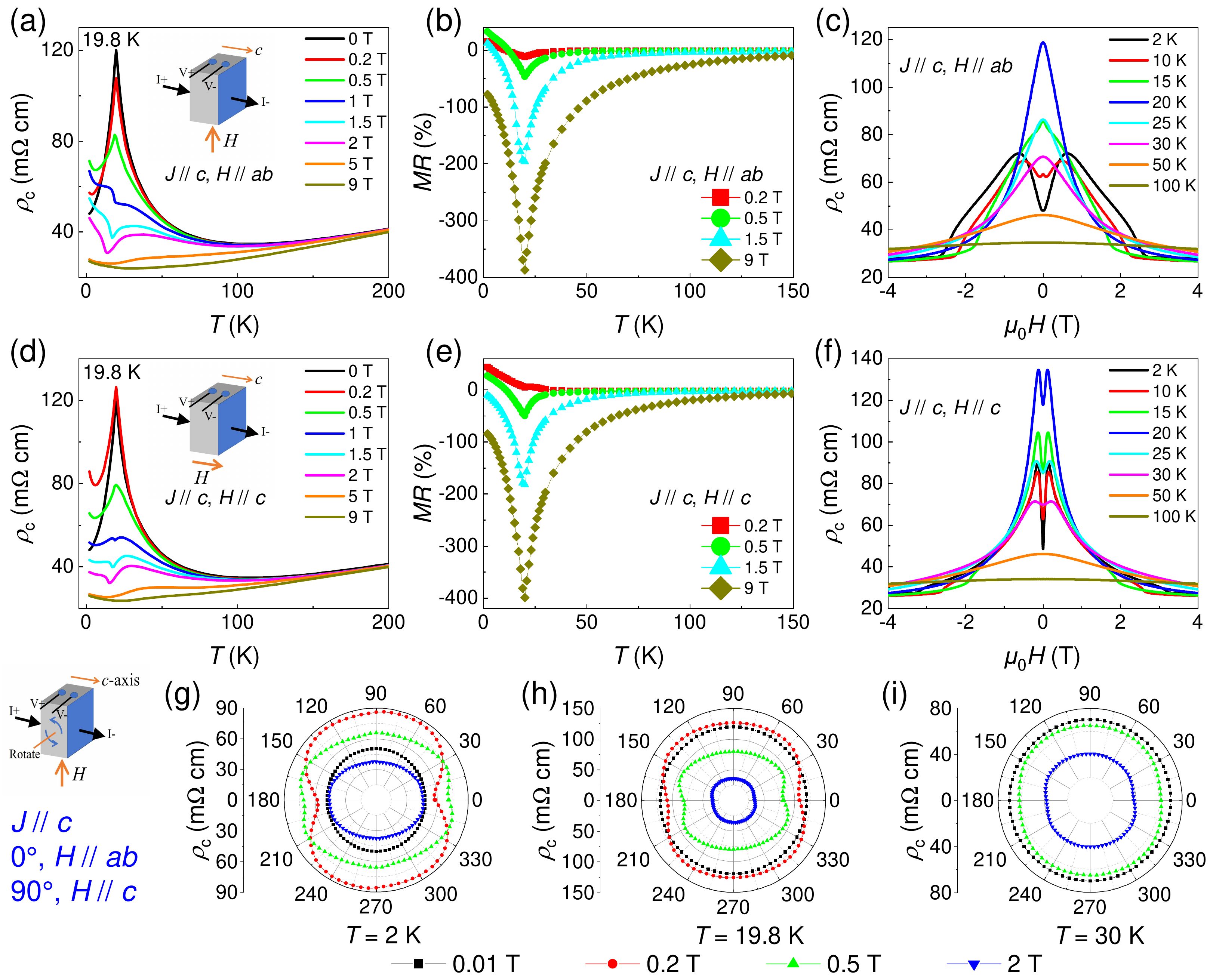}
	\caption{\label{fig:Resistivity-As-c}
		(a) Temperature dependence of out-of-plane resistivity with fields along the $ab$-plane. (b) 
		Magnetoresistance with in-plane fields as a function of temperature. (c) Field dependence of out-of-plane resistivity at several temperatures. (d)(e)(f) $\rho_c(T)$, MR$(T)$ and $\rho_c(H)$ with out-of-plane fields, respectively. (g)(h)(i) Angle dependence of out-of-plane resistivity under several fields at 2 K, 19.8 K and 30 K, respectively. The behavior of $\rho_\mathrm{c}$ is close to $\rho_\mathrm{ab}$ shown in the main text. As we mentioned, the resistivity of \EZA~does not strongly depend on the direction of the current. 
	}
\end{figure*}
Figure \ref{fig:Resistivity-As-c} shows the temperature, ﬁeld, and angle dependence of out-of-plane resistivity. As we pointed out in the main text, the behavior of $\rho_c$ is very close to that of $\rho_{ab}$, i.e., the resistivity for \EZA~is nearly independent of the direction of the current.


\section{Magnetization data}
\begin{figure}[h]
	\centering
	\includegraphics[width=0.5\columnwidth]{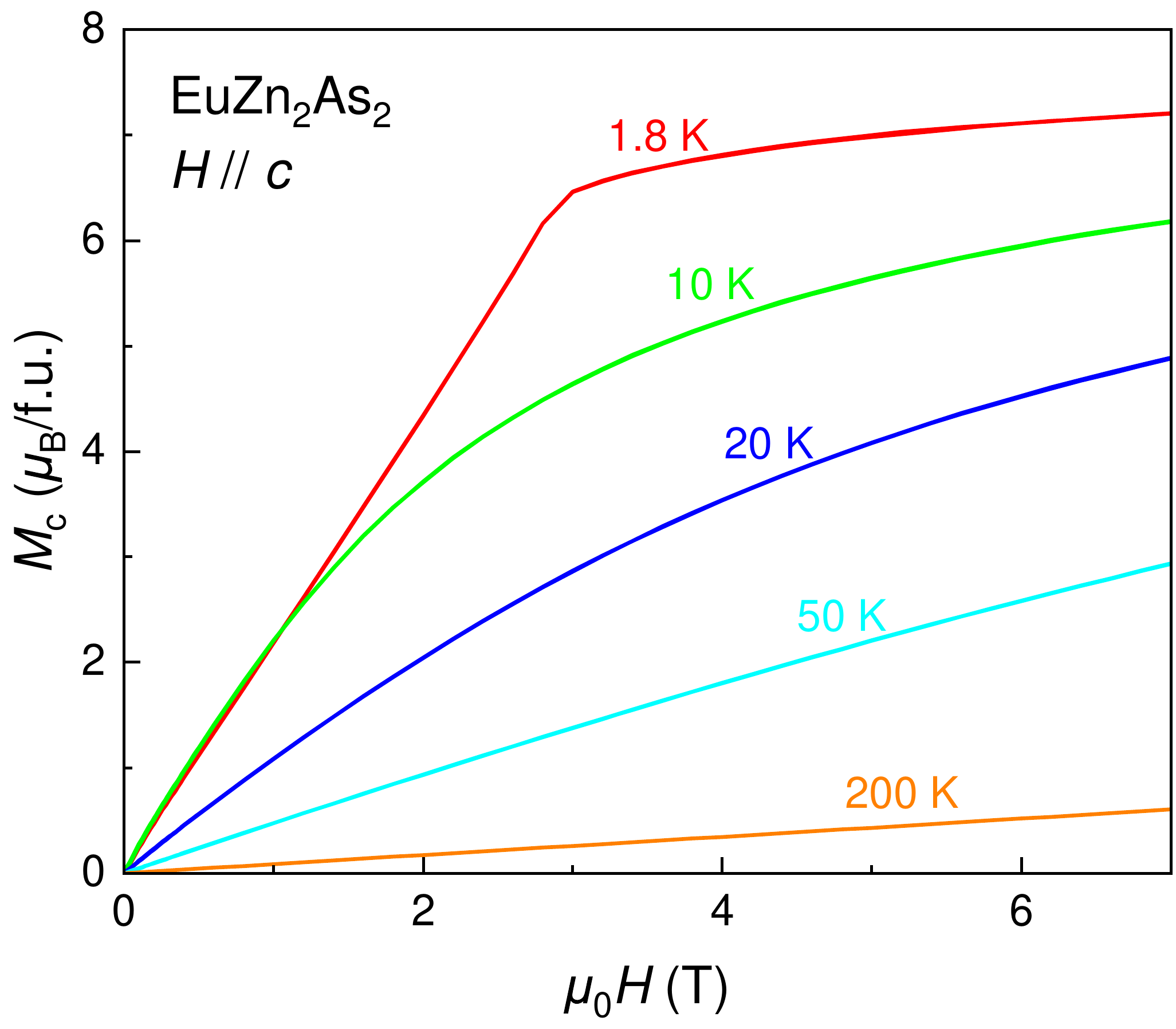}
	\caption{Magnetization as a function of field ($H\|c$) at several temperatures.}
	\label{fig:MH}
\end{figure}
To extract the non-linear AHE, we measured magnetization curves (Fig.~\ref{fig:MH}) at the same temperatures as the Hall resistivity curves (Fig.~5b in the main text). 
As mentioned in the main text, the total Hall resistivity is expressed as $\rho_{xy}=R_0B+R_SM+\rho_{xy}^\mathrm{NA}$. 
The OHE contribution $R_0B$ is linear in $B$ and the linear AHE contribution $R_SM$ is proportional to $M$. 
The non-linear AHE could be obtained after subtracting the OHE and AHE contributions $R_0B+R_SM$. The magnetization is saturated or varies mildly at low temperatures when the field is high. 
So we can take a constant value for $M$, then $R_0B+R_SM$ is only dependent on $B$. 
We fit the original Hall resistivity $\rho_{xy}(H)$ ($T$ = 20 K) from 5 T to 7 T to the equation $R_0B+R_SM$, where $R_0$ is the resulting slope. $R_S$ is calculated from the intercept $R_SM$. 
Here we take $M=6\mu_\mathrm{B}$ for the Hall data ($T$ = 20 K) when $B>$ 5 T. 
With $R_0$, $R_S$ and $M(H)$ data at 20 K, the linear contributions $R_0B+R_SM$ can be calculated and subtracted from the total Hall data.

\section{Details about first-principles calculations}
\label{sec:appendix-theory}

\subsection{Magnetic Structure}
\label{subsec:dft_magneticStructure}

All of the magnetic configurations that were considered for the DFT study of magnetic ground state are summarized in Table \ref{tab:graphLabels}.

\begin{table}[h]
	\caption{\label{tab:graphLabels}  List of all magnetic configurations considered in the DFT calculations (Figs.~\ref{fig:DFTplusDipole}, S4, S5, and S6). The asterisks label configurations that rotate the spin by 120$^{\circ}$ along the $a$ and $b$ directions. Details of the lattice supercells are shown in Figure \ref{fig:ball-stick}}
	\begin{ruledtabular}
		\begin{tabular}{p{1cm}p{13.5cm}}			
			Label & Magnetic configuration \\
			\hline
			f\_b   &  Ferromagnetic (FM), spins aligned along $b$ axis \\
			f\_c  & FM, spins aligned along $c$ axis  \\
			aAb*   &   Type-A antiferromagnetic (AFM), spins aligned in $ab$ plane with 120$^{\circ}$ rotations \\
			aAb     &   Type-A AFM, spins aligned along $b$ axis  \\
			aAc &  Type-A AFM, spins aligned along $c$ axis   \\
			aCb*    &  Type-C AFM, spins aligned in $ab$ plane with 120$^{\circ}$ rotations \\
			aCb  &  Type-C AFM, spins aligned along $b$ axis  \\
			aCc  &  Type-C AFM, spins aligned along $c$ axis  \\
			aGb    & Type-G AFM, spins aligned along $b$ axis \\
			aGc   &  Type-G AFM, spins aligned along $c$ axis \\
		\end{tabular}
	\end{ruledtabular}
\end{table}

\begin{figure}
	\centering
	\includegraphics[width=\columnwidth]{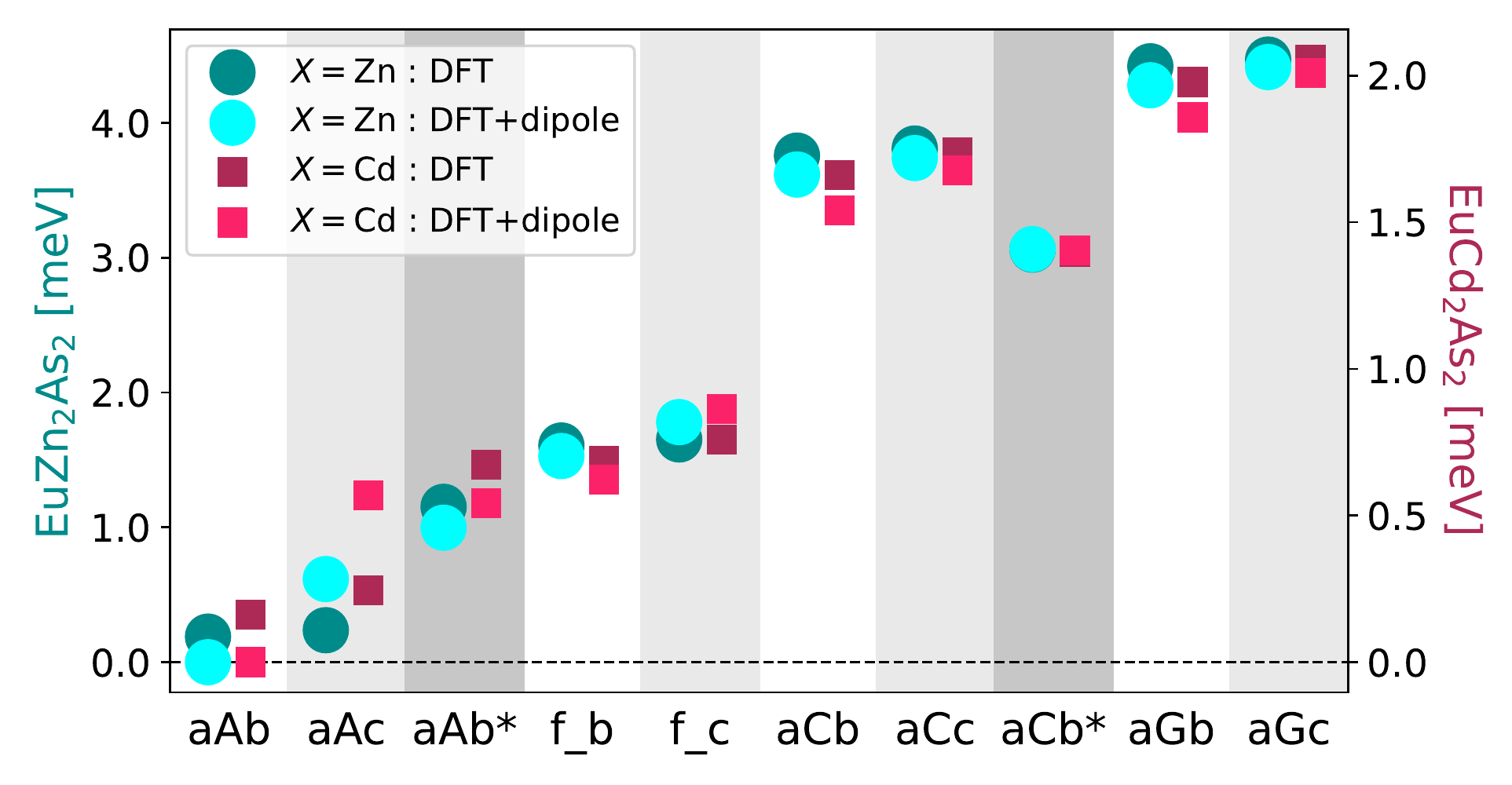}
	\caption{Total energies per europium for EuX$_2$As$_2$ (X= Zn, Cd) for different magnetic configurations, referenced from the ground state magnetic configuration. The ground state for both materials is predicted to be type-A AFM order, however it is notable that the scale of energy between magnetic configurations is quite small, around 5 meV for \EZA\ and only around 2 meV for \ECA.}
	\label{fig:DFTplusDipole}
\end{figure}

The total energies for each magnetic configuration, calculated from the full GGA+$U$+SOC scheme of DFT as described at the end of Section II, is plotted in Fig.~\ref{fig:DFTplusDipole}. We also added a dipole coupling correction to the DFT total energy by calculating the classical magnetic dipole-dipole interactions between the large magnetic moments of Eu$^{2+}$. The dipole coupling correction results in a slight shift but not a change of preferred ground-state. Defects and interstitials likely have a larger effect on total energy than the dipole coupling, but this is left to future work.

\begin{figure}
	\centering
	\includegraphics[width=0.7\columnwidth]{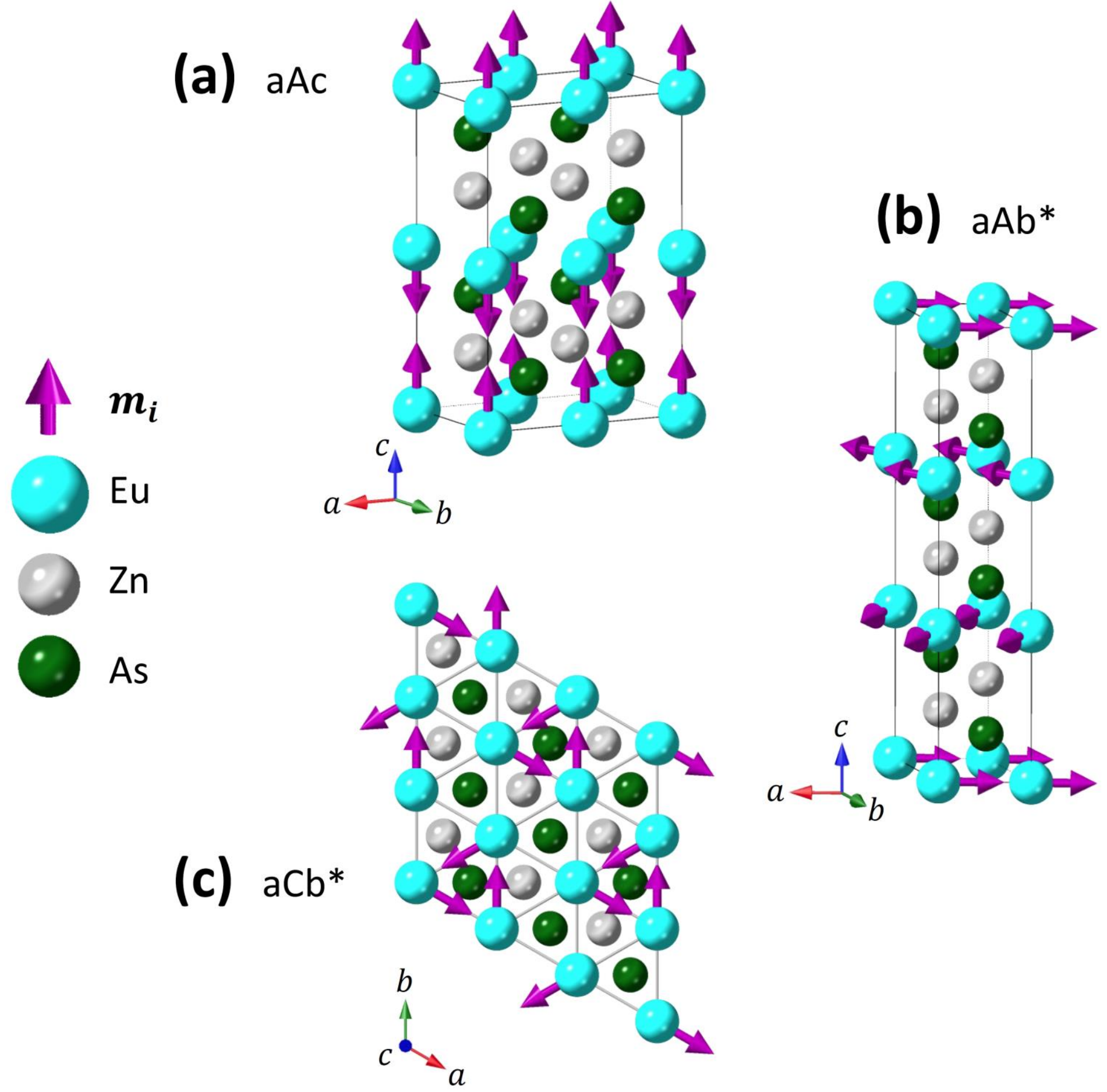}
	\caption{Diagrams of the lattice configurations and supercells for different magnetic orders. Panel (a) shows the supercell used for every calculation except the 120$^{\circ}$ spin rotation configurations. These 120$^{\circ}$ spin rotation versions of type-A AFM and type-C AFM are shown in panels (b) and (c), respectively.}
	\label{fig:ball-stick}
\end{figure}

To calculate the various magnetic configurations, large enough lattice supercells were constructed to capture the symmetry of the magnetic ordering.  In order to make the most meaningful comparisons, all of the calculations were performed in the same large supercell to capture the magnetic order variations, with two exceptions. Diagrams of the supercells are shown in Figure \ref{fig:ball-stick}. 

All calculations were done in the 4-Eu supercell shown in panel (a) except for the two configurations which incorporated 120$^{\circ}$ rotations of spins aligned in the $a$-$b$ plane (labeled with asterisks). 
The first of these variations, the type-A AFM (aAb*), used a 3-Eu supercell which tripled the unit cell in the $c$ axis direction, as shown in panel (b). 
The second of these variations, the type-C AFM (aCb*), used a 9-Eu supercell which tripled the unit cell along both the $a$ and $b$ axes, as shown in panel (c).

\begin{figure*}[h]
	\centering
	\includegraphics[width=\textwidth]{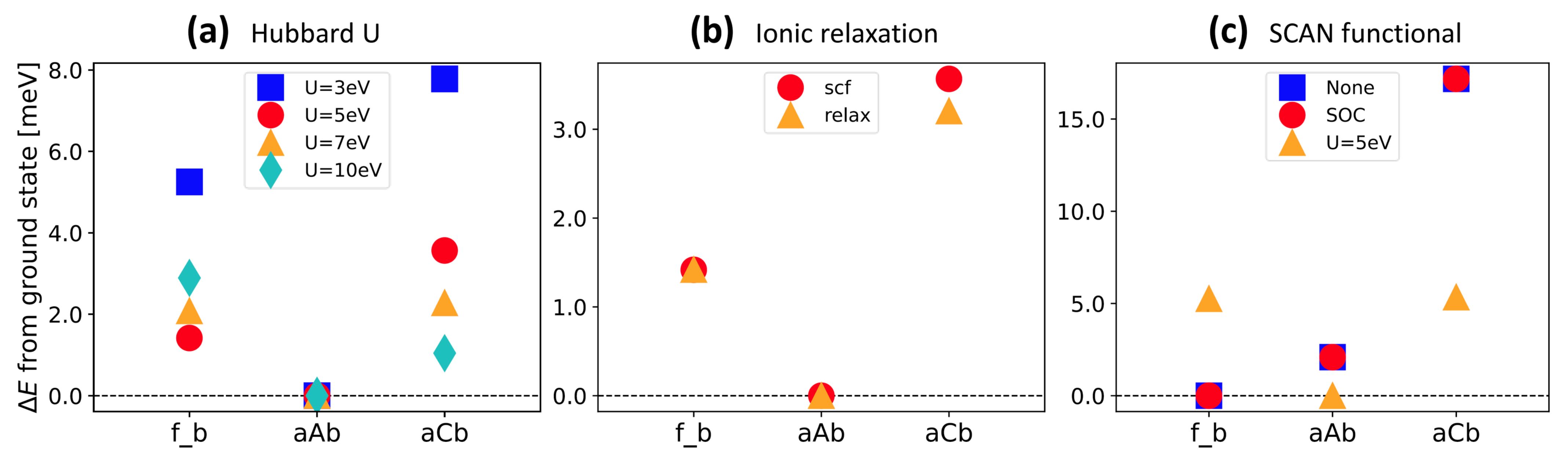}
	\caption{DFT adjustments for EuZn$_2$As$_2$ to determine ground state magnetic configuration.}
	\label{fig:DFTchanges}
\end{figure*}

The DFT calculations were attempted with different parameters to discern the reliability of the method in determining total energy of magnetic ground states (Figure \ref{fig:DFTchanges}). 
This analysis was only performed for EuZn$_2$As$_2$. There was minimal difference in energy between spins aligned along $b$ or $c$, so only $b$ is shown. 
The type-G AFM always showed the highest energy, and has been excluded from figures. 

As shown in Figure \ref{fig:DFTchanges} panel (a), the value of the Hubbard $U$ on the Eu $4f$ electrons was varied to find the effect on the magnetic ground-state. 
It is clear that type-A AFM always has the lowest energy, but it is interesting to see that the excited state switches between FM and type-C AFM as $U$ increases. 
The highest $U$ considered here is equivalent to previous first principles calculations on EuZn$_2$As$_2$ \cite{EuZn2Sb2_Uvalue}. 
In panel (b), we show the difference in total energies when the experimental structure was used versus when the ions were in their DFT-determined relaxed positions. 
We performed unrestricted variable-cell relaxation calculations for each magnetic configuration and report the final total energy. 
In panel (c) we attempted to use a more sophisticated Meta-GGA functional \cite{SCAN_performance}, SCAN (Strongly constrained and appropriately normed semilocal density functional) \cite{SCAN_original}.
When there is no Hubbard $U$ using the SCAN potential, the ferromagnetic configuration is the ground state, regardless of spin-orbit-coupling. 
However, when $U=5$~eV, the clear ground state is AFM type-A. The results shown in Figure \ref{fig:DFTchanges} further support that the ground state magnetic configuration of EuZn$_2$As$_2$ is type-A AFM, while also supporting the conclusion that the magnetic configurations are very close in energy (within a few meV) and intrinsic effects like disorder or extrinsic influences such as strain or other applied fields may be able to tip the balance between competing states quite easily.

\subsection{Electronic Structures}

\begin{figure}[h]
	\centering
	\includegraphics[width=\textwidth]{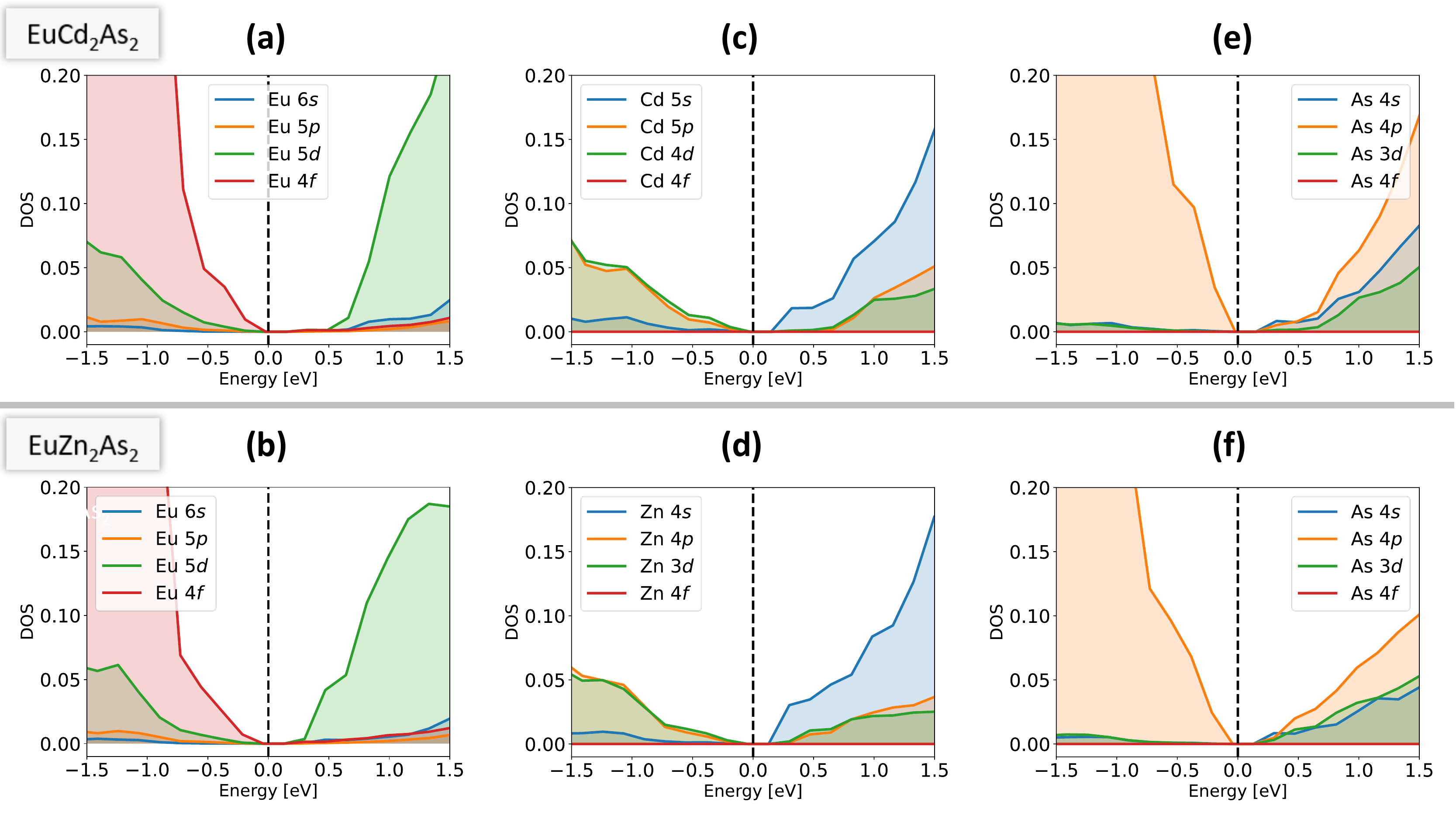}
	\caption{The orbitally resolved density of states for \ECA and \EZA. Subplots \textbf{(a, b)} show the orbitals of Eu, \textbf{(c, d)} the orbitals of Cd and Zn, respectively, and \textbf{(e, f)} the orbitals of As.}
	\label{fig:ldos}
\end{figure}

The orbitally projected density of states is plotted in Fig.~\ref{fig:ldos}. The dominant character of the conduction band in both cases is the highest $s$ orbital of $X$ in Eu$X_2$As$_2$. There are a great deal of similarities between \ECA\ and \EZA\ from these plots. The nuances of the differences between materials become more apparent when looking at the momentum-resolved aspects of the band structure.

\begin{figure*}[h]
	\centering
	\includegraphics[width=\columnwidth]{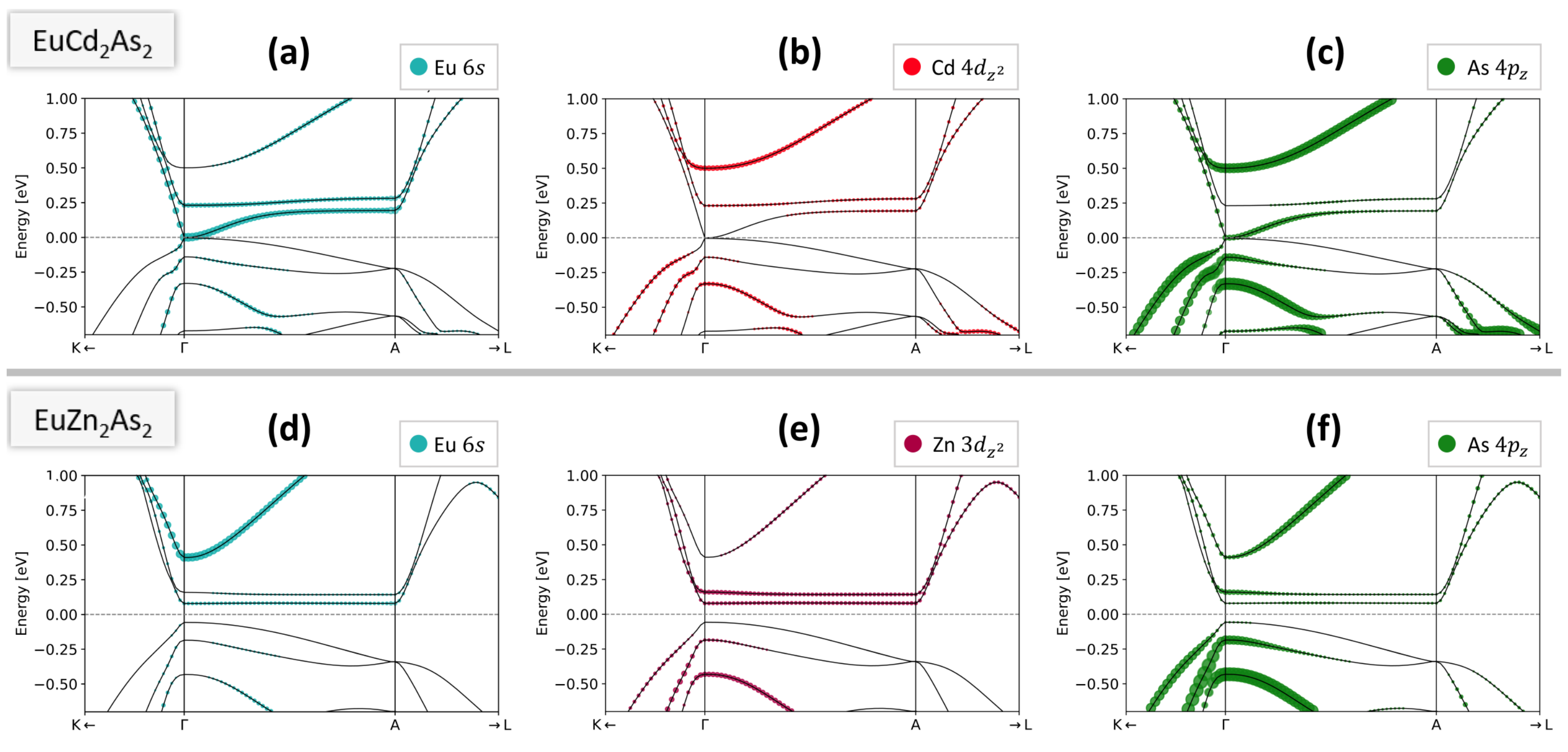}
	\caption{Band projections into the Eu-$s$ \textbf{(a, b) }, Cd- and Zn-$d$ \textbf{(c, d) }, and As-$p$ orbitals in \textbf{(e, f)} for \ECA\ and \EZA.}
	\label{fig:kpdos}
\end{figure*}

We plotted the orbital projections onto the band structures of \ECA\ and \EZA\, and summarize the most interesting differences in Fig.~\ref{fig:kpdos}. The difference in position of the strongest Eu-$6s$ character band is very noticeable between \ECA\ and \EZA. The order of the strongest bands for Cd(Zn) $4(3)d_{z^2}$ and As $4p_z$ orbital projections seem to also switch orders when changing from Cd to Zn in Eu$X_2$As$_2$. Spin-orbit coupling could be the tuning parameter that pushes down the third band above the Fermi energy to the conduction band when going from Zn to Cd.

\bibliography{Wang_17Feb2022_SI}